\documentclass[10pt]{article}
\usepackage[french,english]{babel}
\usepackage{amsmath}
\usepackage{amsfonts}
\usepackage{amssymb}
\usepackage{graphicx}                              % pour figures pdf
\usepackage{times}
\usepackage{version}
\usepackage{framed,color}
\usepackage[makeroom]{cancel}
\begin{document}
%=======================================================================
%
July 7th, 2014   \hfill % CoquandMachet.tex

%\hfill arXiv:
%\today\quad   \hfill %summary3.tex
\vskip 7cm
{\baselineskip 12pt
\begin{center}
{\bf FROM QUANTUM FIELD THEORY TO NANO-OPTICS :

\bigskip
\hbox{\hskip -5mm REFRACTIVE PROPERTIES OF GRAPHENE
                IN A MEDIUM-STRONG EXTERNAL MAGNETIC FIELD}}
\end{center}
}
\baselineskip 16pt
%=====================================================================
\arraycolsep 3pt  %space between columns in matrices
%=====================================================================
%
\vskip .2cm
\centerline{O. Coquand
     \footnote[1]{Ecole Normale Sup\'erieure, 61 avenue du Pr\'esident
Wilson, F-94230 Cachan}
     \footnote[2]{ocoquand@ens-cachan.fr}
, B.~Machet
     \footnote[3]{Sorbonne Universit\'e, UPMC Univ Paris 06, UMR 7589,
LPTHE, F-75005, Paris, France}
     \footnote[4]{CNRS, UMR 7589, LPTHE, F-75005, Paris, France.}
     \footnote[5]{Postal address:
LPTHE tour 13-14, 4\raise 3pt \hbox{\tiny \`eme} \'etage,
          UPMC Univ Paris 06, BP 126, 4 place Jussieu,
          F-75252 Paris Cedex 05 (France)}
    \footnote[6]{machet@lpthe.jussieu.fr}
     }
\vskip 1cm

{\bf Abstract:} 1-loop quantum corrections are shown to induce large effects
on the refraction index $n$ inside a graphene strip
in the presence of an external magnetic field $B$ orthogonal to it.
To this purpose, we use the tools of Quantum Field Theory  to calculate
the photon propagator at 1-loop inside  graphene in position space,
which leads to an effective vacuum polarization in a brane-like theory
of  photons interacting with  massless 
electrons at locations  confined  inside the thin strip  (its
longitudinal spread is considered to be infinite).
The effects factorize into  quantum ones, controlled by the value of
$B$ and that of the electromagnetic coupling $\alpha$, 
and a ``transmittance function'' $U$ in which the geometry of the sample and 
the resulting confinement of electrons  play the major roles.
We consider photons  inside the visible spectrum
and  magnetic fields  in the range 1-20\; Teslas.
At $B=0$, quantum effects depend very weakly
on  $\alpha$ and $n$ is essentially controlled by  $U$; we recover, then,
 an opacity for visible light of the same order of magnitude $\pi \alpha_{vac}$
 as measured experimentally.

%\smallskip

%{\bf PACS:}

%================================================================
\newpage\tableofcontents

\newpage\listoffigures\newpage
%================================================================

%SSSSSSSSSSSSSSSSSSSSSSSSSSSSSSSSSSSSSSSSSS
\section{Introduction}\label{section:intro}
%SSSSSSSSSSSSSSSSSSSSSSSSSSSSSSSSSSSSSSSSSS

Very strong magnetic fields $B$ are known to induce dramatic effects on
the spectrum of hydrogen and on the critical number $Z_c$ of atoms
\cite{MachetVysotsky} \cite{GodunovMachetVysotsky}.
However, the typical effects being ${\cal O}(\frac{e^3 B}{m_e})$, 
 gigantic fields are needed, $\geq 10^{16}$ Gauss, which are out of reach
on earth.

The property that the fine structure constant $\alpha$ in graphene largely exceeds
$1$ \cite{Goerbig} instead of its vacuum value
$\alpha_{vac}\simeq\frac{1}{137}$ was a sufficient motivation
to investigate whether  sizable effects could be obtained at lower cost in
there.

While graphene in an external magnetic field is usually associated to the
so-called ``abnormal quantum hall effect'' \cite{Goerbig} \cite{CastroNeto},
 our results show that one can
also expect optical effects for the visible spectrum and ``reasonable''
values of the magnetic field.

This work relies on the Schwinger formalism \cite{Schwinger}
\cite{DittrichReuter}
 to write the propagator of the
Dirac-like massless electrons inside graphene 
in an external magnetic field $B$ perpendicular to the graphene strip,
 and on a calculation in position space of
the photon propagator at 1-loop. This enables  in particular to
explicitly constrain the vertices of electrons with photons to
stay confined inside the graphene strip
\footnote{Sometimes we write abusively about the confinement of electrons,
but we always mean that the vertices at which they interact with photons
lie inside graphene and are therefore confined in the $z$ direction
between $-a$ and $+a$.}. So doing, we get an
effective 1-loop vacuum polarization $\Pi^{\mu\nu}_{eff}$
 that can be plugged in the light-cone
equations derived according to the pioneering work of Tsai and Erber
\cite{TsaiErber1}, and of Dittrich and Gies \cite{DittrichGies1}.

One of the salient features of the effective $\Pi^{\mu\nu}_{eff}$ is that
it factorizes into the 1-loop $\Pi^{\mu\nu}$ calculated with the standard
rules of Quantum Field Theory (QFT) in the presence of an external magnetic
field, adapted of course to the case of the Hamiltonian for graphene at the
Dirac points, times a universal function $U$ which does not depend on the
magnetic field, but only on the energy $q_0$ of the photon, of the relative
penetration $u$ inside the graphene strip (very weakly), and of its ``geometry'' (in a
somewhat extended meaning). It is classical by nature and shares
similarities with the so called ``transmittance'' function in optics or
``transfer function'' in electronics.
The genuine $\Pi^{\mu\nu}$  concentrates
all quantum effects and those of the magnetic field.

It is also a remarkable fact that, though electrons inside graphene
correspond classically to massless $3+1$ Dirac electrons with
a vanishing momentum $p_3$ along the direction of $B$, the quantum
calculation of the photon propagator at 1-loop shows that the latter can
nevertheless exchange momentum with virtual electrons in the direction of $B$ as
expected from quantum mechanics and their confinement inside the small
width $2a$ of the graphene strip:  the corresponding
unavoidable ``energy-momentum non-conservation'' of photons
 along $B$ is found indeed to be the quantum uncertainty $\hbar/a$ on the
electron momentum.

The effects that we describe only concern photons with ``parallel''
polarization (see \cite{TsaiErber1}) (for transverse polarization,  the
only solution that we found to the light-cone equation is the trivial $n=1$).
The large value of the electromagnetic coupling turns out not to be 
only amplifying factor. That the  massless Dirac electrons of graphene that
interact with photons are confined  inside a thin strip plays also  a major
 role.

The effects of the confinement of electrons that arise add
to the ones that have, for example,  been investigated  in
 \cite{Perez} when the longitudinal size $L$ of graphene is finite.
 Confinement conspires with the external magnetic field
 to produce macroscopic effects. The
difference is that we are concerned here with the confinement in the
``short'' direction, the thickness $2a\approx 350\, pm$
 of graphene, considering that its
large direction $L$ is like infinite
\footnote{The cyclotron radius $\ell = \sqrt{\frac{\hbar}{eB}}$ is $\ell
\approx 8.1\,10^{-9}\, m$ at $B=10\,T$.}.
This makes an intuitive physical
interpretation much less easy since, now, no cyclotron radius can
eventually reach the (longitudinal)  size of the graphene strip.

The refractive index $n=n_1+i\,n_2$
 is found to essentially depend on $\alpha$, on the angle of
incidence $\theta$, and on the ratio
$\Upsilon=\frac{\sqrt{2eB}}{q_0}$.
In the absence of any external $B$, its dependence 
on the electromagnetic coupling fades away, and it is mainly constrained by
the sole property that electrons are confined.

A transition occurs at small angle of incidence $\theta_{min} \sim
\frac{1}{\Upsilon}$: no solution with $|n_2| \ll n_1$ to the light-cone
equation exists anymore for $\theta < \theta_{min}$. There are hints that
the system goes brutally to a regime with large index/absorption. The
identification of the corresponding solutions however requires more
elaborate numerical techniques, which will be the object of a subsequent
work.

That the final results differ from what would be obtained from $QED_{2+1}$
is expected since the gauge field, unlike the electrons, lives in
$3+1$ dimension and the framework of our calculations is more
a ``brane-like'' picture for graphene.

Our calculations are made in the limit of a ``medium-strong'' $B$,
in the sense that $\sqrt{2eB} \gg q_0$,
 and are only valid at this limit such that, in particular, the case $B= 0$
is not directly accessible. This is why we have performed a special calculation
 for $B=0$ the end of the study. $B$ is however not considered to be ``infinite''
like in \cite{MachetVysotsky} \cite{GodunovMachetVysotsky} \cite{Godunov}.
They also require $a q_0 \ll 1$, in which $2a$ is the thickness of the graphene
strip. This last approximation guarantees to stay in the linear part of the
electron spectrum close to the Dirac point, which is an essential
ingredient to use a ``Dirac-like'' effective Hamiltonian \cite{Goerbig}.
We are concerned with photons in the visible spectrum, which sets us very
far from geometrical optics
\footnote{The corresponding wavelengths are indeed roughly 3 orders of
magnitude larger that the thickness of graphene.}, and limit $B$, for the sake 
of experimental feasibility, to 20 Teslas.

Our results are summarized in the 2 plots of Figure\;\ref{fig:nBreal}.

The last section  is dedicated to the case when no external
magnetic field is present. We show in this case that no $\theta_{min}$
exists and that, instead, when the angle of incidence gets smaller and
smaller, the refraction index $n$  goes continuously from ``quasi-real'' values to
complex values with larger $n_1$ and $n_2$. At very small values of
$\theta$, we recover an opacity of the same order
of magnitude  as the one measured experimentally \cite{Nair}.

The literature dedicated to graphene is enormous and we cannot
unfortunately pay  a tribute to the whole of it. We only make few
citations, but the reader can find, in particular inside the reviews
articles, references to most of the important works.

%SSSSSSSSSSSSSSSSSSSSSSSSSSSSSSSSSSSSSSSSSSSSSSSSSSSSSSSSSSSSSSSSSSSS
\section{From the vacuum polarization to light-cone equations and to
the refraction index}\label{section:general}
%SSSSSSSSSSSSSSSSSSSSSSSSSSSSSSSSSSSSSSSSSSSSSSSSSSSSSSSSSSSSSSSSSSSS

\subsection{Conventions and setting}\label{subsec:settings}
%=========================================================

\begin{figure}[h]
\begin{center}
\includegraphics[width=10 cm, height=7 cm]{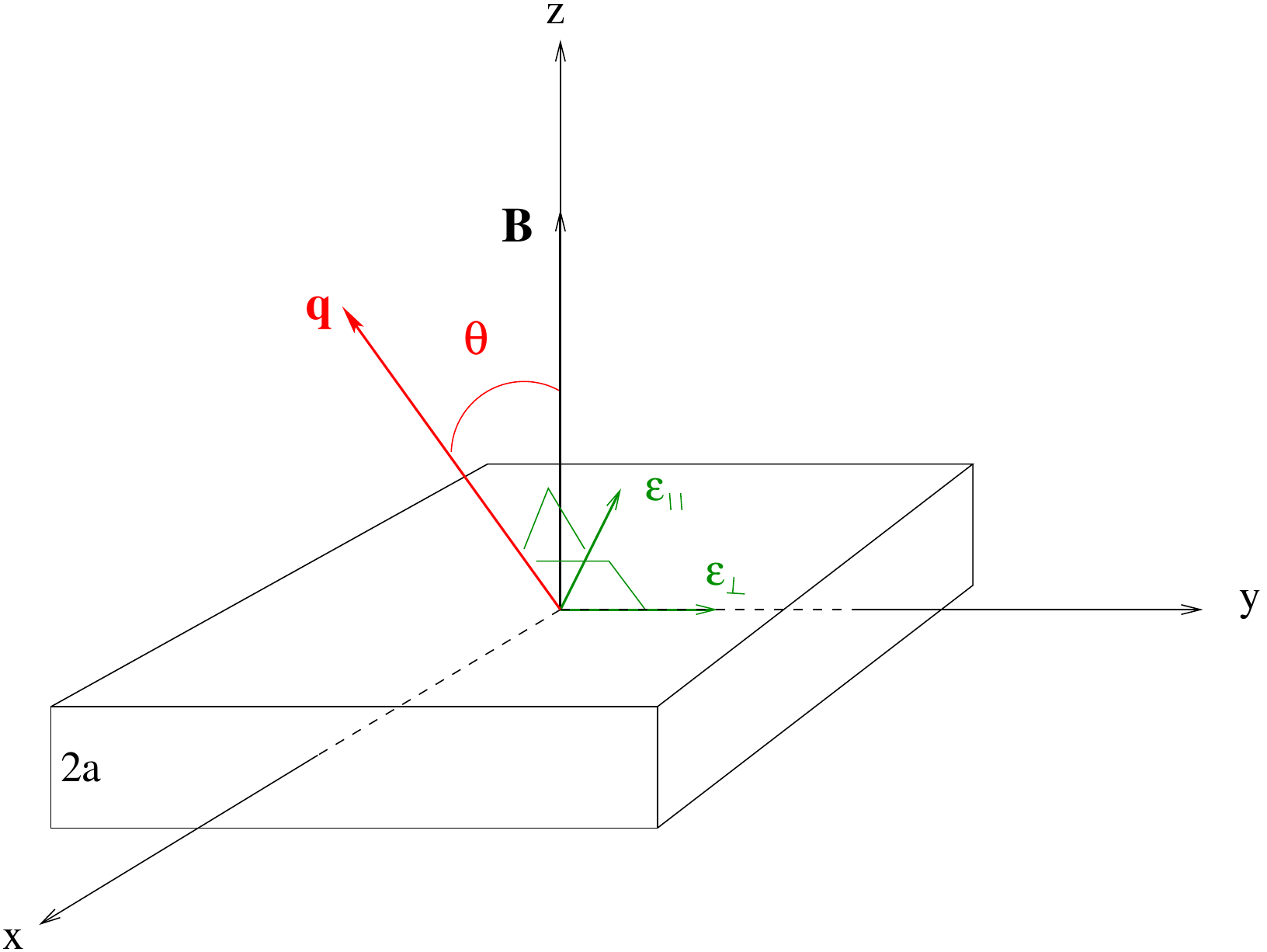}
\caption{$\vec B$ is perpendicular to the graphene strip of width $2a$.
The polarization vector $\vec\epsilon$, perpendicular to the momentum $\vec
q$ of the electromagnetic wave, is decomposed into $\vec\epsilon_\parallel $ in
the $(x,z)$ plane and $\vec\epsilon_\perp$ perpendicular to this
plane.} \label{fig:setup}
\end{center}
\end{figure}

Let us follow Tsai-Erber \cite{TsaiErber1}. $\vec B$ is
 $\parallel z$, the $\vec q$ of the propagating photon (plane wave)
is chosen to lie in the $(x,z)$ plane. See Fig.~\ref{fig:setup}.

We shall call $\theta$ the ``angle of incidence''; the reader should
keep in mind that, since we are concerned with the propagation of light
{\em inside} graphene, $\theta$  is  the angle of incidence of light
{\em inside this medium}.

The polarization vector $\vec\epsilon$\; is decomposed
into $\vec\epsilon_\parallel$ and $\vec\epsilon_\perp$, both orthogonal to
$\vec q$. $\vec\epsilon_\perp \parallel y$ and $\vec\epsilon_\parallel$
is in the $(x,z)$ plane. If we call $\theta$
 the angle $(\vec B, \vec q)$,
 $\vec\epsilon_\parallel = -\cos\theta\, \vec i + \sin\theta\, \vec k$,
while $\vec\epsilon_\perp=\vec j$.
One has $\vec q= |\vec q|\,(\sin\theta\,\vec i + \cos\theta\,\vec
k)$.
We shall call $\vec\epsilon_\parallel$ ``parallel polarization'' and
$\vec\epsilon_\perp$ ``transverse polarization''.
It must be noticed that, at normal incidence $\theta=0$, there is no longer
a plane $(\vec q, B)$ such that these 2 polarizations can no longer be
distinguished.

We shall in the following use ``hatted'' letters for vectors living in the
Lorentz subspace $(0,1,2)$. For example
\begin{equation}
\hat q = (q^0,q^1,q^2),\quad q=(\hat q, q_3) = (q_0,q_1,q_2,q_3).
\end{equation}

\subsection{The modified Maxwell Lagrangian and the light-cone equations}
\label{subsec:lightcone}
%=======================================================================

Taking into account the contribution of
 vacuum polarization, the Maxwell Lagrangian gets modified to
\cite{DittrichGies1}
\begin{equation}
{\cal L}(x)= -\frac14 F_{\mu\nu}(x)F^{\mu\nu}(x) -\frac12 \int d^4 y\;
A^\mu(x)\Pi_{\mu\nu}(x, y,B) A^\nu( y),
\label{eq:maxL}
\end{equation}
from which one gets the Euler-Lagrange equation
\begin{equation}
\Big( g_{\mu\nu}\,q^2 -q_\mu q_\nu + \Pi_{\mu\nu}(q, B)\Big)
A^\nu(q)=0.
\label{eq:maxeq}
\end{equation}
Left-multiplying (\ref{eq:maxeq}) with
\begin{equation}
A^\mu= \alpha\epsilon^\mu_\perp + \beta\epsilon^\mu_\parallel,
\label{eq:Apol}
\end{equation}
yields
\footnote{When $\Pi_{\mu\nu}$ is not present,  the only non-vanishing
elements are ``diagonal'',
$\epsilon^\mu_\perp \Big( g_{\mu\nu}\,q^2 -q_\mu
q_\nu\Big)\epsilon^\nu_\perp=q^2
=\epsilon^\mu_\parallel \Big( g_{\mu\nu}\,q^2 -q_\mu
q_\nu\Big)\epsilon^\nu_\parallel$, which yields
$A^\mu \Big( g_{\mu\nu}\,q^2 -q_\mu
q_\nu\Big) A^\nu = (\alpha^2 + \beta^2) q^2$, and, accordingly,
the customary light-cone condition $q^2=0 \equiv q_0^2 -\vec q^2$.
If $\Pi_{\mu\nu}$ is transverse $\Pi_{\mu\nu}= (g_{\mu\nu}q^2
-q_\mu q_\nu)\Pi(q^2)$, the light-cone condition is $(\alpha^2+\beta^2)q^2
(1+\Pi(q^2))=0$, that is, as usual, $q^2=0$. }

\begin{equation}
\begin{split}
& (\alpha\epsilon^\mu_\perp + \beta\epsilon^\mu_\parallel)
\Big( g_{\mu\nu}\,q^2 -q_\mu q_\nu + \Pi_{\mu\nu}(q,  B)\Big)
(\alpha\epsilon^\nu_\perp + \beta\epsilon^\nu_\parallel)=0,\cr
& \epsilon^\mu_\perp = (0,0,1,0),\quad
\epsilon^\mu_\parallel=(0,-c_\theta,0,s_\theta),\qquad
c_\theta\equiv\cos\theta,\ s_\theta\equiv\sin\theta.
\end{split}
\end{equation}
We shall identify $\Pi^{\mu\nu}$ with the {\em effective polarization} inside
graphene $\Pi^{\mu\nu}_{eff}$ that we shall calculate  in section
\ref{section:propagator} and therefore consider in the following, instead
of (\ref{eq:maxeq}), the Euler-Lagrange equation
\begin{equation}
\Big( g^{\mu\nu}\,q^2 -q^\mu q^\nu + \Pi^{\mu\nu}_{eff}(q, B)\Big)
A_\nu(q)=0.
\label{eq:maxeq2}
\end{equation}
As we shall see there,
$\Pi_{eff}^{03}=0=\Pi_{eff}^{13}=\Pi_{eff}^{23}$, such that we shall be
concerned with the simplified light-cone equation
\begin{equation}
(\alpha^2 + \beta^2)q^2 + \left(\alpha^2 \Pi_{eff}^{22}(q,B) + \beta^2 \left(
c_\theta^2\;
\Pi_{eff}^{11}(q,B) +
s_\theta^2\; \Pi_{eff}^{33}(q,B)\right) + 2\alpha\beta\,c_\theta\,
\Pi_{eff}^{12}(q,B)\right)
=0.
\label{eq:lcgen}
\end{equation}
$\vec q$ has been furthermore chosen to lie in the $(x,z)$ plane, so
$q_2=0$, which entails
(see (\ref{eq:summary})) $\Pi^{02}_{eff}=0=\Pi_{eff}^{20}, \Pi^{12}_{eff}=0 =
\Pi_{eff}^{21}$,
 and the light-cone relation finally shrinks to
\begin{equation}
(\alpha^2 + \beta^2)q^2 + \left(\alpha^2 \Pi_{eff}^{22}(q,B) + \beta^2 \left(
c_\theta^2\;
\Pi_{eff}^{11}(q,B) + s_\theta^2\; \Pi_{eff}^{33}(q,B)\right)\right)
=0.
\label{eq:lc0}
\end{equation}

Depending of the polarization of the photon, there are accordingly
2 different light-cone  relations:\newline
$\bullet$\ for $A^\mu_\perp(q_0,q_1,0,q_3)$, $\alpha=1,\beta=0$,

\begin{equation}
q^2 + \Pi_{eff}^{22}(q,B) =0;
\label{eq:lcperp}
\end{equation}
$\bullet$\  for  $A^\mu_\parallel(q_0,q_1,0,q_3)$, $\alpha=0,\beta=1$,
\begin{equation}
q^2 + \Big(c_\theta^2\; \Pi_{eff}^{11}(q,B) + s_\theta^2\;
\Pi_{eff}^{33}(q,B)\Big) =0.
\label{eq:lcpar}
\end{equation}
One of the main features of (\ref{eq:lcpar}) is the occurrence of
$\Pi^{33}_{eff}$, which would not be there in $QED_{2+1}$. We shall see
later that this term plays an important role.

A remark is due concerning eq.~(\ref{eq:maxeq}). It derivation from the
effective Lagrangian (\ref{eq:maxL}) relies on the property that
$\Pi_{\mu\nu}(x,y)$ is in reality a function of $(x-y)$ only. This is however
not the case for $\Pi^{\mu\nu}_{eff}$ which, as we shall see, depends
indeed on $(\hat x -\hat y)$ but individually on $x_3$ and $y_3$ (see the
first remark at the end of subsection \ref{subsub:graph}). Once the
dependence on $(x_3-y_3)$ has been extracted, there is a left-over dependence
on $y_3$, which finally yields for our results the dependence of the
refraction index on $u=\frac{y_3}{a} \in[-1,+1]$. We shall see however that
this dependence is always extremely weak, and we consider therefore 
the Euler-Lagrange equation (\ref{eq:maxeq2})  to be valid to a very good approximation.

\subsection{The refractive index}\label{subsec:index}
%====================================================

We define it in a standard way by
\begin{equation}
n= \frac{|\vec q|}{q_0}.
\label{eq:ndef}
\end{equation}

In practice, $\Pi^{\mu\nu}_{eff}$ is not only a function of $q$ and $B$,
but of the angle of incidence $\theta$ and of the relative depth $u$ inside
the graphene strip, $u\in [-1, +1] $.
The light-cone equations therefore translate into relations
$n=n(\theta,B,q_0,u)$ that we will write explicitly after calculating
$\Pi^{\mu\nu}$ and $\Pi^{\mu\nu}_{eff}$.

%SSSSSSSSSSSSSSSSSSSSSSSSSSSSSSSSSSSSSSSSSSSSSSSSSSSSSSSSSSSSSSSSSSSS
\section{Calculation of the 1-loop vacuum polarization
$\boldsymbol{\Pi^{\mu\nu}}(\hat q,B)$
 in the presence of an external magnetic field $\boldsymbol B$}
\label{section:vacpol}
%SSSSSSSSSSSSSSSSSSSSSSSSSSSSSSSSSSSSSSSSSSSSSSSSSSSSSSSSSSSSSSSSSSSS

\begin{figure}[h]
\begin{center}
\includegraphics[width=6 cm, height=3 cm]{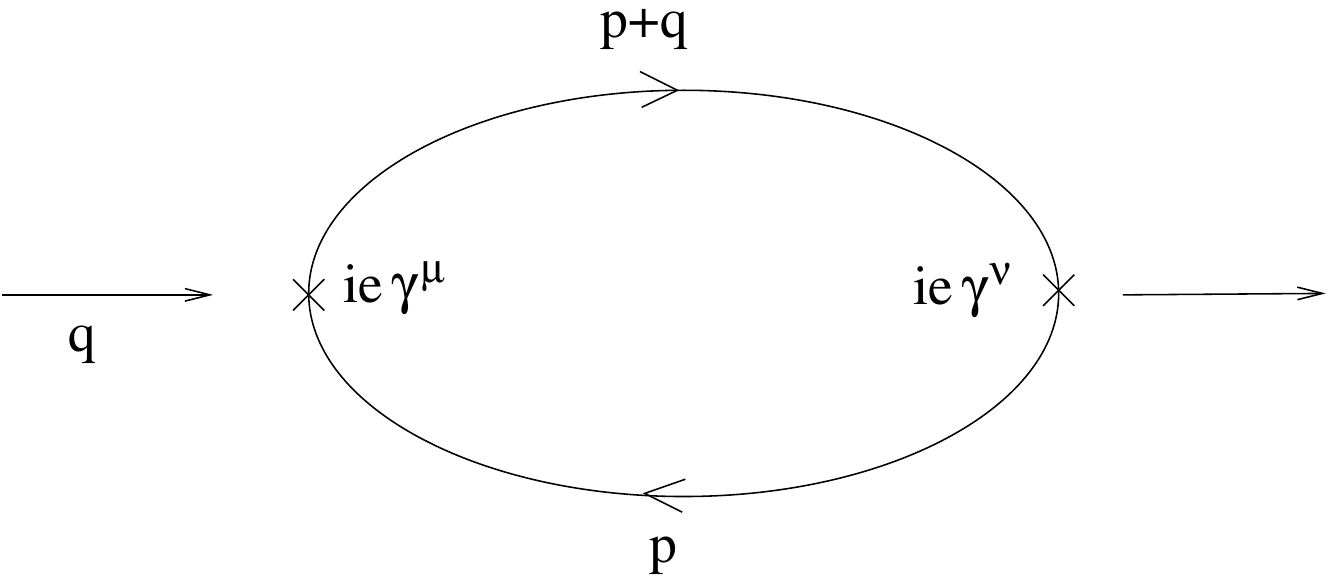}
\caption{The vacuum polarization $\Pi^{\mu\nu}$}
\label{fig:vacpol}
\end{center}
\end{figure}

It is given by
\begin{equation}
i\Pi^{\mu\nu}(\hat q,B)= +e^2\int_{-\infty}^{+\infty} \frac{d^3\hat
p}{(2\pi)^3}\; Tr\; \left[\gamma^\mu G(\hat p,B)
\gamma^\nu G(\hat p+\hat q,B)\right],
\label{eq:vapog}
\end{equation}
in which $G(\hat p,B)$ is the  propagator of a massless Dirac electron
obtained from the Hamiltonian of graphene at the Dirac points,  making use
of the formalism of Schwinger \cite{Schwinger}\cite{Tsai1974-2}
 to account for the external magnetic field $B$.

\subsection{The electron propagator in an external magnetic field}
\label{subsec:eprop}
%=================================================================

Following Schwinger ((\cite{Schwinger}, eqs.\, 2.7 to 2.10),
 we define the electron propagator as
%(note the $i$)
\begin{equation}
G(x,y) = i\langle(\psi(x) \bar\psi(y))_+\rangle \Theta(x-y).
\end{equation}

The graphene at the Dirac points is described (see for example
\cite{Goerbig} \cite{CastroNeto})
 by a Hamiltonian which is exactly Dirac in 4 dimensions
but with $m_e=0$ and $p_3=0$ (and the $\gamma$ matrices 
taken in the chiral representation).

The propagator of an electron inside graphene will accordingly be taken to
be \cite{Schwinger}\cite{Tsai1974-2}
\footnote{The expression (\ref{eq:eprop1}) is obtained after going from the
real proper-time $s$ of Schwinger to $\tau=is$ and switching to conventions
for the Dirac matrices and for the metric of space $(+,-,-,-)$
which are more usual today \cite{PeskinSchroeder}.
%However \cite{TsaiErber1}, the change of coutour $s
%\to is$ is only allowed in the range of photon frequencies 
%below the pair creation theshhold $q_0 < 2m_e$.
}
\begin{equation}
\hskip -1cm  G(\hat p,B)=
\int_0^{\infty} d\tau\; \exp\left[
-\tau\left((-p_0^2)+ \frac{\tanh (e\tau B)}{e\tau B}(p_1^2+p_2^2)\right)
\right]
\left((\gamma^0 p^0)(1-i\gamma^1\gamma^2 \tanh(e\tau B))-\frac{\gamma_1
p_1 + \gamma_2 p_2}{\cosh^2(e\tau B)}\right),
\label{eq:eprop1}
\end{equation}
which only depends on $\hat p=(p_0, p_1, p_2)$ and $B$.

\subsubsection{Expanding at ``large'' $\boldsymbol{B<\infty}$}
\label{subsub:bigB}
%-------------------------------------------------------------

At the limit $B\to \infty$
\footnote{One considers then that $e\tau B$ also $\to \infty$, in which case,
in
(\ref{eq:eprop1})
$\tanh e\tau B \to 1, \cosh e\tau B \to \infty$.
This is only acceptable at $\tau \not=0$, but Schwinger's prescription is that
the integration over the proper time has to be made last.}
, (\ref{eq:eprop1}) becomes
\begin{equation}
G(\hat p,B)\stackrel{B=\infty}{\to} -e^{-\frac{p_\perp^2}{eB}}\;
\frac{\gamma^0 p^0}{p_0^2}(1- i\gamma^1\gamma^2),\quad
p_\perp^2 = p_1^2+p_2^2.
\label{eq:Ginf}
\end{equation}

We shall in this work go one step further in the expansion of $G$ at large
$B$: we keep the first subleading terms in the expansions of $\tanh (\tau eB)$ and
$\cosh (\tau eB)$ of (\ref{eq:eprop1})
\footnote{This approximation does not allow later to take the limit $B\to
0$ since, for example, it yields $\tanh (\tau eB)
\to -1$
instead of $0$ and $\cosh^2(\tau eB) \to 3/4$ instead of $1$.}
:
\begin{equation}
\tanh (\tau eB) \approx 1-2 e^{-2\tau eB},\quad \cosh^2(\tau eB) \approx
\frac{e^{2\tau eB}+2}{4} \Rightarrow \frac{1}{\cosh^2(\tau eB)} \approx
\frac{4\,e^{-2\tau eB}}{1+2\,e^{-2\tau eB}}.
\end{equation}

This gives (we note $ (\gamma p)_\perp=\gamma_1p_1+\gamma_2p_2$), still for
graphene,
\begin{equation}
\begin{split}
G(\hat p,B)\approx &\int_0^\infty d\tau \;
e^{-\tau (-p_0^2)}e^{-\frac{p_\perp^2}{eB}(1-2e^{-2e\tau B})}
(\gamma^0p^0)(1-i\gamma^1\gamma^2(1-2e^{-2e\tau B}))\cr
   &- 4 (\gamma p)_\perp \int_0^\infty d\tau \;
e^{-2e\tau B}\frac{1}{1+2e^{-2e\tau B}}e^{-\tau(-p_0^2)}e^{-\frac{p_\perp^2}{eB}
(1+2e^{-2e\tau B})}.
\end{split}
\end{equation}
We shall further approximate
 $e^{-\frac{p_\perp^2}{eB}(1-2e^{-2e\tau B})} \approx
e^{-\frac{p_\perp^2}{eB}}$, which can be seen to be legitimate  because the
exact integration yields subleading corrections $\propto 1/(eB)^2$,
while the ones that we keep are $\propto 1/eB$. This gives
\begin{equation}
\begin{split}
G(\hat p,B)\approx &
e^{-\frac{p_\perp^2}{eB}}\left(-\frac{\gamma^0p^0}{p_0^2}
(1-i\gamma^1\gamma^2)+2
\frac{\gamma^0p^0}{p_0^2-2eB}(-i\gamma^1\gamma^2)\right) 
-4 (\gamma p)_\perp e^{-\frac{p_\perp^2}{eB}}\int_0^\infty d\tau \;
\frac{1}{1+2e^{-2e\tau B}}e^{-\tau (-p_0^2+2eB)}.
\end{split}
\label{eq:gfinexact}
\end{equation}

One has
\begin{equation}
\int_0^\infty d\tau \;
\frac{1}{1+2e^{-2e\tau B}}e^{-\tau (-p_0^2+2eB)}=
(-2)^{-1+\frac{p_0^2}{2eB}}
\frac{\beta(-2,1-\frac{p_0^2}{2eB},0)}{2eB},
\label{eq:genint}
\end{equation}
such that  (\ref{eq:gfinexact}) rewrites
\begin{equation}
\hskip -.5cm
G(\hat p,B)
 = -e^{-\frac{p_\perp^2}{eB}} \left(
\frac{\gamma^0}{p^0}\left(1+i\gamma_1\gamma_2\;
\frac{p_0^2+2eB}{p_0^2-2eB}\right)
+4\frac{p_1\gamma_1+p_2\gamma_2}{2eB} F(\frac{p_0^2}{2eB})
\right),\quad F(x)=(-2)^{(-1+x)}\beta(-2,1-x,0),
\label{eq:gfing2}
\end{equation}
in which $\beta$ is the incomplete beta function.

When $B < \infty$, corrections arise with respect to (\ref{eq:Ginf}),
 which exhibit in particular poles at
$p_0^2=2eB$ (first and 2nd term) and also $p_0^2 = 2n\,eB, n=1,2 \ldots$
 (second term)
\footnote{If we do not work explicitly for graphene, one finds that the
electron mass squared
 $m_e^2$ gets replaced by $m_e^2 + 2n\,eB$ in the presence of $B$.}
.

\subsubsection{Our working approximation}\label{subsub:Fappro}
%-------------------------------------------------------------

The expression (\ref{eq:gfing2}) is still not very simple to use. This is
why we shall  further approximate $F(x)$ and take
\begin{equation}
F(x) \approx \frac{1}{1-x},
\label{eq:workapp}
\end{equation}
which amounts to only select, in there, the pole at $p_0^2 = 2eB$ and
neglect the other poles.
This approximation is reasonable in the vicinity of this pole,
 for $0 \leq x \leq 1.5$, that is $0 \leq p_0^2 \leq 1.5\times (2eB)$, low
energy (massless) electrons. It will be discussed more in subsection
\ref{subsec:lowen} in which we show that the approximation is valid for
electrons with energy $p_0 \leq 10\,eV$ at the weakest magnetic fields
$B=1\,T$ that we consider..

We shall accordingly take
\begin{equation}
\begin{split}
G(\hat p,B) &\approx
-e^{-\frac{p_1^2+p_2^2}{eB}} \left[
\frac{\gamma^0}{p^0}\left(1+i\gamma_1\gamma_2\;
\frac{p_0^2+2eB}{p_0^2-2eB}\right)
-4\frac{p_1\gamma_1+p_2\gamma_2}{p_0^2-2eB}
\right].
\end{split}
\label{eq:ggap}
\end{equation}
This leads to  expressions much  easier to handle, and enables to go a long way
analytically.

\subsection{Calculation and results}
%==================================

There are 2 steps in the calculation. First one has to perform the traces
on the Dirac $\gamma$ matrices, then do explicitly the integration over
the loop variable $\hat p$.

\subsubsection{Performing the traces of $\boldsymbol \gamma$ matrices}
\label{subsub:technique}
%---------------------------------------------------------------------

This step already yields
\begin{equation}
\Pi^{i3}=0=\Pi^{3i},\quad i=0,1,2.
\label{eq:Pii3}
\end{equation}

\subsubsection{Doing the integrations} \label{subsub:integrate}
%-------------------------------------------------------------

\begin{figure}[h]
\begin{center}
\includegraphics[width=5 cm, height=3 cm]{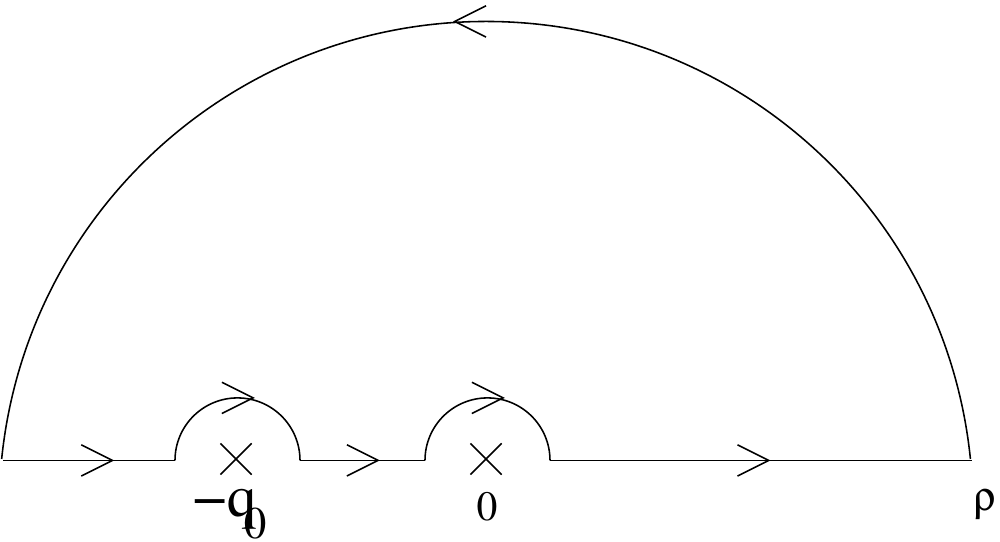}
\caption{The contour of integration for $B(q_0)$ and $C(q_0)$}
\label{fig:contour}
\end{center}
\end{figure}

Details of the calculation will be given somewhere else.
We just want here to present its main steps, taking the example of
$\Pi^{00}$. After doing the traces, one gets
\begin{equation}
\begin{split}
i\Pi^{00}(\hat q,B) &=
4\,e^2\int_{-\infty}^{+\infty} \frac{dp_0 dp_1 dp_2}{(2\pi)^3}
e^{-p_\perp^2/eB}e^{-(p+q)_\perp^2/eB}\cr
& \left(\frac{1}{p_0}\frac{1}{p_0+q_0}
+\frac{1}{p_0}\frac{p_0^2+2eB}{p_0^2-2eB}
\frac{1}{p_0+q_0}\frac{(p_0+q_0)^2 +2eB}{(p_0+q_0)^2-2eB}
+16 \frac{p_1(p_1+q_1)+p_2(p_2+q_2)}{(p_0^2-2eB)((p_0+q_0)^2-2eB)}
\right),
\end{split}
\label{eq:pi00}
\end{equation}
which decomposes into
\begin{equation}
\begin{split}
i\Pi^{00}(\hat q,B) &= I(\hat q,B) + J(\hat q,B) + K(\hat q,B), \cr
I(\hat q,B) &= 4\,e^2\int_{-\infty}^{+\infty} \frac{dp_0 dp_1 dp_2}{(2\pi)^3}
e^{-p_\perp^2/eB}e^{-(p+q)_\perp^2/eB}
 \frac{1}{p_0}\frac{1}{p_0+q_0}, \cr
J(\hat q,B) &= 4\,e^2\int_{-\infty}^{+\infty} \frac{dp_0 dp_1 dp_2}{(2\pi)^3}
e^{-p_\perp^2/eB}e^{-(p+q)_\perp^2/eB}
\frac{1}{p_0}\frac{p_0^2+2eB}{p_0^2-2eB}
\frac{1}{p_0+q_0}\frac{(p_0+q_0)^2 +2eB}{(p_0+q_0)^2-2eB}, \cr
K(\hat q,B) &= 4\,e^2\int_{-\infty}^{+\infty} \frac{dp_0 dp_1 dp_2}{(2\pi)^3}
e^{-p_\perp^2/eB}e^{-(p+q)_\perp^2/eB}
16 \frac{p_1(p_1+q_1)+p_2(p_2+q_2)}{(p_0^2-2eB)((p_0+q_0)^2-2eB)}.
\end{split}
\label{eq:pi00c}
\end{equation}
It is then convenient to introduce
\begin{equation}
\begin{split}
B(q_0)&= \int_{-\infty}^{+\infty} dp_0 \frac{1}{p_0}\frac{1}{p_0+q_0},\cr
C(q_0,B) &= \int_{-\infty}^{+\infty} dp_0
\frac{1}{p_0}\frac{p_0^2+2eB}{p_0^2-2eB}
\frac{1}{p_0+q_0}\frac{(p_0+q_0)^2 +2eB}{(p_0+q_0)^2-2eB},\cr
D(q_0,B)&= \int_{-\infty}^{+\infty} dp_0
 \frac{1}{(p_0^2-2eB)((p_0+q_0)^2-2eB)},
\end{split}
\label{eq:Bdef}
\end{equation}
such that, integrating over the transverse degrees of freedom $p_1,p_2$,
one gets
\begin{equation}
\begin{split}
I(\hat q, B) &= \frac{\alpha}{\pi}eB\;e^{-q_\perp^2/2eB} B(q_0),\cr
J(\hat q, B) &= \frac{\alpha}{\pi}eB\;e^{-q_\perp^2/2eB} C(q_0,B),\cr
K(\hat q, B) &= \frac{8\alpha}{\pi}eB\;e^{-q_\perp^2/2eB}
(eB-q_\perp^2) D(q_0,B).
\end{split}
\end{equation}
``Massless'' and ambiguous integrals of the type
 $\int_{-\infty}^{+\infty} d\sigma \;\frac{f(\sigma)}{\sigma}$ occurring in
$B(q_0), C(q_0,B), D(q_0,B)$ are
replaced, using the customary $+i\varepsilon$ prescription for the poles of
propagators in QFT dictated by causality, with
\begin{equation}
\lim _{\epsilon\to 0^+} \int_{-\infty}^{+\infty} d\sigma
\;\frac{f(\sigma)}{\sigma+i\epsilon} = -i\pi\,f(0)
+\lim_{\epsilon \to 0^+} \int_{|\sigma| >\epsilon}
\frac{f(\sigma)}{\sigma},
\end{equation}
which are just Cauchy integrals. This is nothing more than the
Sokhotski-Plemelj theorem \cite{SokhotskiPlemelj} :
\begin{equation}
\lim_{\varepsilon\to 0^+} \int_{-\infty}^\infty\frac{f(x)}{x\pm
i\varepsilon}\,dx = \mp i\pi f(0) + \lim_{\varepsilon\to 0^+}
\int_{|x|>\varepsilon}\frac{f(x)}{x}\,dx.
\end{equation}
It is easy to also check that the same result can be obtained, after
setting the $+i\varepsilon$ prescription, by integrating on the contour
described on Fig.\ref{fig:contour}. There, the 2 small 1/2 circles around the poles have
radii that $\to 0$. The large 1/2 circle has infinite radius.

This also amounts, for the poles ``on the real axis'',
 to evaluating $i\pi \sum residues$, that is $1/2$ of
what one would get if the poles were not on the real axis but inside the
contour of integration. The other poles that lie inside the contour of
integration are dealt with as usual by $2i\pi \times$ their residues.

So doing, one gets
\begin{equation}
\begin{split}
& B(q_0) =0 = C(q_0,B),\cr
& D(q_0,B)= 2i\pi
(-)\frac{1}{\sqrt{2eB}}\frac{1}{q_0^2-8eB},
\end{split}
\end{equation}
leading finally to 
\begin{equation}
I=0=J,\quad  K(\hat q,B)=
i \frac{2e^2}{\pi}\;e^{-\frac{q_\perp^2}{2eB}}
\sqrt{2eB}\frac{eB -  q_\perp^2}{q_0^2-8eB},
\end{equation}
and, for $\Pi^{00}(\hat q, B)$, to the first line of the set of equations
 (\ref{eq:summary}).

After all integrals have been calculated by this technique, one gets the
following results.

\subsubsection{Explicit expression of the vacuum polarization at
1-loop}\label{subsub:cavpol}
%---------------------------------------------------------------

\begin{equation}
\begin{split}
i\Pi^{00}(\hat q,B) &= 4i\alpha\;\sqrt{2eB}\;e^{-\frac{q_1^2+q_2^2}{2eB}}\;
\frac{2eB -2(q_1^2+q_2^2)}{q_0^2 -4(2eB)} \stackrel{B\to\infty}{\simeq}
-i\alpha\;\sqrt{2eB}\;e^{-\frac{q_1^2+q_2^2}{2eB}},\cr
i\Pi^{11}(\hat q,B) &= 4i\alpha\,
e^{-\frac{q_1^2+q_2^2}{2eB}}\sqrt{2eB}\,\frac{q_1^2-q_2^2}{q_0^2-4(2eB)}
\stackrel{B\to\infty}{\simeq}
i\alpha\,e^{-\frac{q_1^2+q_2^2}{2eB}}\,\frac{q_1^2-q_2^2}{\sqrt{2eB}}, \cr
i\Pi^{22}(\hat q,B) &= -i\Pi^{11}(\hat q,B),\cr
i\Pi^{01}(\hat q,B) &= 2i\alpha\, e^{-\frac{q_1^2+q_2^2}{2eB}} q_1q_0\,
\frac{\sqrt{2eB}}{q_0^2-2eB} \stackrel{B\to\infty}{\simeq}
-i\alpha\,e^{-\frac{q_1^2+q_2^2}{2eB}}\,\frac{q_1q_0}{\sqrt{2eB}},\cr
i\Pi^{02}(\hat q,B) &=   2i\alpha\, e^{-\frac{q_1^2+q_2^2}{2eB}} q_2q_0\,
\frac{\sqrt{2eB}}{q_0^2-2eB} \stackrel{B\to\infty}{\simeq}
-i\alpha\,e^{-\frac{q_1^2+q_2^2}{2eB}}\,\frac{q_2q_0}{\sqrt{2eB}},\cr
i\Pi^{12}(\hat q,B) &= -16\alpha\, q_1q_2\, e^{-\frac{q_1^2+q_2^2}{2eB}}
\frac{\sqrt{2eB}}{q_0^2-4(2eB)} \stackrel{B\to\infty}{\simeq}
4\alpha\,e^{-\frac{q_1^2+q_2^2}{2eB}}\, \frac{q_1q_2}{\sqrt{2eB}},\cr
i\Pi^{33}(\hat q,B) &= -i\Pi^{00}(\hat q,B)
%\stackrel{B\to\infty}{\simeq} i\alpha\;\sqrt{2eB}\;e^{-\frac{q_1^2+q_2^2}{2eB}}
,\cr
i\Pi^{03}(\hat q,B)&=0,\quad i\Pi^{13}(\hat q,B)=0,\quad i\Pi^{23}(\hat q,B)=0.
\end{split}
\label{eq:summary}
\end{equation}

\subsubsection{Comments}\label{subsub:comments}
%----------------------------------------------

$\bullet$\ $\Pi^{00}=-\Pi^{33}$ are the only 2 components that
do not vanish when $B\to\infty$ nor when $\theta \to 0$.

$\bullet$\  The formula (22) of  \cite{Godunov} taken at $m_e=0$ yields
$\Pi^{33}=-\Pi^{00}$; we get the same relation;
it also yields $\Pi^{03} \propto Tr ((k-q)_3 k_0 + (k-q)_0 k_3)$ such
that, at $k_3=0=(k-q)_3$ it  yields $0$ like we get.

$\bullet$\ Transversality is broken since  we do not have $q_\mu
\Pi^{\mu\nu}(q)=0$. At the opposite the
general formula (34) in Tsai-Erber \cite{TsaiErber1}
for the vacuum polarization in magnetic field
is shown in their eq. (36) to satisfy gauge invariance.

In \cite{Godunov} the transversality conditions reduce to
$q^0 \Pi_{03}+ q^3 \Pi_{33}=0= q^0\Pi_{00}+ q^3\Pi_{30}$, the other
relations being automatically satisfied. Now, as can be easily checked in
there, at  $k_3=0=(k-q)_3$,
$\Pi_{03}=0=\Pi_{30}$, $\Pi_{33}\not=0$, $\Pi_{00}\not=0$, and the
transversality conditions, which reduce to $q^3 \Pi_{33}=0=q^0\Pi_{00}$
 can non longer be satisfied either (unless $q_3=0=q_0$).
So, restraining the electrons to have a
vanishing momentum along the direction of $B$ breaks ``gauge invariance''.

%SSSSSSSSSSSSSSSSSSSSSSSSSSSSSSSSSSSSSSSSSSSSSSSSSSSSSSSSSSSSSSSSSSSSSSSS
\section{The photon propagator in $\boldsymbol x$-space and the effective
vacuum polarization $\boldsymbol{\Pi_{eff}^{\mu\nu}}$}
\label{section:propagator}
%SSSSSSSSSSSSSSSSSSSSSSSSSSSSSSSSSSSSSSSSSSSSSSSSSSSSSSSSSSSSSSSSSSSSSSSS

The vacuum polarization that needs to be introduced inside the light-cone
equations (\ref{eq:lcperp},\ref{eq:lcpar}) 
is not $\Pi^{\mu\nu}(\hat q, B)$ computed in section \ref{section:vacpol}
 above, but the effective
 $\Pi^{\mu\nu}_{eff}(\hat q, q_3,\frac{y_3}{a}, B)$ obtained by
calculating the photon propagator in position-space, while confining, at the 2
vertices $\gamma\, e^+ e^-$, the corresponding $z$'s to lie inside graphene,
$z\in [-a,a]$ ($2a$ is the graphene width).

The effective polarization writes
$\Pi^{\mu\nu}_{eff}(\hat
q,q_3,\frac{y_3}{a},B)=\frac{1}{\pi^2}\Pi^{\mu\nu}(\hat q,B)\;
U(\hat q, q_3, \frac{y_3}{a})$
in which  $U$ is a universal function that does
not depend on the magnetic field and that we also encounter when dealing
with the case of no external $B$. It is the Fourier transform of the
product of 2 functions:  the first, $\frac{\sin a k_3}{ak_3}$,  is the
Fourier transform of the ``gate function'' corresponding to
the graphene strip along
$z$; the second  carries  the remaining
information attached to the confinement of electrons.
Its analytical properties inside the complex plane control in particular
the ``leading'' $\frac{1}{\sin\theta}$ behavior of the refraction index
inside graphene, where $\theta$ is the angle of incidence inside the
graphene strip (see subsection \ref{subsec:settings}).
The integration variable of this transformation is $k_3$, the
difference between the momenta along $B$ of
the outgoing and incoming photons (see below).

This factorization can be traced back to the fact
that $\Pi^{\mu\nu}$ does not depend on
$q_3$, for the simple reason that the Hamiltonian of electrons at
the Dirac points inside graphene has $p_3=0$.
An example of how factors combine is the following.
$\Pi^{\mu\nu}_{eff}$ still includes an integration on $p_3$ (the 
component along $B$ of the momentum of the virtual electron inside graphene).
Like $\Pi^{\mu\nu}(\hat q, B)$, this integral factors out.
Since electrons are confined along $B$, $p_3$ cannot, quantum-mechanically,
exceed $\pm\frac{1}{a}$ such  the integral becomes simply  proportional to
$\frac{1}{a}$. This factor  completes, inside the integral $\int dk_3$ defining
 $U$, the ``geometric'' $\frac{\sin ak_3}{ak_3}$ evoked above.

$k_3$ represents the amount of energy-momentum non-conservation of photons
along $B$: this phenomenon cannot indeed but occur  at vertices between
3+1-dimensional photons and ``confined'' electrons (like, as we already
mentioned, the non-transversality of $\Pi_{\mu\nu}$).
However, the integration $dk_3$  gets 
automatically bounded by the rapid decrease of $\frac{\sin ak_3}{ak_3}$ at
$|k_3| > \frac{1}{a}$  and
this bound is the same as the one for the electron momentum $p_3$,
$|k_3|\leq \frac{1}{a}$. So, the energy-momentum non-conservation
between  the outgoing and incoming photons cannot exceed
the uncertainty on the electron momentum
due to its confinement. In particular, when the graphene strip becomes
infinitely thick $a \to \infty$, this cut-off goes to $\infty$ and one recovers
standard QFT in 3+1 dimensions, with the integration on $k_3$ going from
$-\infty$ to $+\infty$.

\subsection{The 1-loop photon propagator in position space}
\label{subsec:gammaprop}
%==========================================================

We calculate the 1-loop photon propagator
\begin{equation}
 \Delta^{\rho\sigma}(x,y)= \langle 0 \vert T A^\rho(x) A^\sigma(y)\vert
0\rangle
\end{equation}
and somewhat lighten  the notations, omitting symbols like T-product, 
\ldots

Introducing the coordinates $u=(u_0,u_1,u_2,u_3)$ and $v=(v_0,v_1,v_2,v_3)$
of the two $(\gamma\,e^+ e^-)$ vertices one gets at 1-loop
\begin{equation}
 \Delta^{\rho\sigma}(x,y)=
 \int d^4u \int d^4v\; A^\rho(x) [(ie) A^\mu(u) \bar\psi(u)\gamma_\mu
\psi(u)]
[(ie)A^\nu(v) \bar\psi(v)\gamma_\nu \psi(v)] A^\sigma(y).
\end{equation}
Making the contractions for fermions etc \ldots yields,
\begin{equation}
\begin{split}
 \Delta^{\rho\sigma}(x,y) &= e^2\int d^4u \int d^4v\; Tr
\int \frac{d^4q}{(2\pi)^4}\; e^{iq(u-x)}\Delta^{\rho\mu}(q)
\gamma_\mu \int \frac{d^4p}{(2\pi)^4}\; e^{ip(u-v)}G(p)
\gamma_\nu\cr
& \hskip 4cm \int \frac{d^4r}{(2\pi)^4}\; e^{ir(v-u)}G(r)
\int \frac{d^4s}{(2\pi)^4}\; e^{is(y-v)}\Delta^{\sigma\nu}(s).
\end{split}
\end{equation}
In what follow we shall always omit writing the trace symbol ``$Tr$''.

\subsubsection{Standard QFT}
%---------------------------

 One integrates $\int_{-\infty}^{+\infty}
d^4u$ and $\int_{-\infty}^{+\infty} d^4v$  for the 4 components of $u$ and
$v$. This gives:
\begin{equation}
 \Delta^{\rho\sigma}(x,y)=\int \frac{d^4q}{(2\pi)^4}\; e^{-iq(x-y)}
\Delta^{\rho\mu}(q)
\Delta^{\nu\sigma}(q)
\underbrace{e^2\int \frac{d^4p}{(2\pi)^4}\; \gamma_\mu G(p) \gamma_\nu
G(p+q)}_{i\Pi_{\mu\nu}(q)}.
\end{equation}
When calculating the vacuum polarization the 2 external photon propagators
have to be removed, which gives
\begin{equation}
 i\Pi_{\mu\nu}(q)=+e^2\int \frac{d^4p}{(2\pi)^4}\; \gamma_\mu G(p)
\gamma_\nu G(p+q).
\end{equation}

\subsubsection{The case of graphene electrons confined along $\boldsymbol z$}
\label{subsub:graph}
%---------------------------------------------------------------------------

The coordinates $u_3$ and $v_3$ of the 2 vertices we do not integrate anymore
$\int_{-\infty}^{+\infty}$ but only $\int_{-a}^{+a}$ in which $2a$ is the
thickness of the graphene strip. This restriction {\em localizes the
interactions of electrons  with photons inside graphene}.

So doing, the result that we shall get will only be valid inside graphene,
and we shall therefore focus on the ``optical properties'' of graphene.
Indeed, photons also interact with electrons outside graphene but, there,
the electron propagators are the ones in the vacuum, not in graphene.

Decomposing $du = d^3\hat u\, du_3, dv=d^3\hat v\, dv_3$, we get by
standard manipulations
\begin{equation}
\begin{split}
\Delta^{\rho\sigma}(x,y)
&=
e^2\int \frac{dp_3}{2\pi} \int \frac{dq_3}{2\pi}\int \frac{dr_3}{2\pi}
\int \frac{ds_3}{2\pi}\int_{-a}^{+a}du_3\; e^{iu_3(q_3+p_3-r_3)}
\int_{-a}^{+a}dv_3\; e^{iv_3(-p_3+r_3-s_3)}
\cr
& \int \frac{d^3\hat q}{(2\pi)^3}\;
e^{i\hat q(\hat y -\hat x)}
e^{iq_3(-x_3)} e^{is_3(y_3)}\Delta^{\rho\mu}(\hat q,q_3)
\Delta^{\sigma\nu}(\hat q, s_3)\
  \int \frac{d^3\hat p}{(2\pi)^3}\;\gamma_\mu G(\hat p) \gamma_\nu G(\hat
p+\hat q).
\end{split}
\end{equation}
Now,
\begin{equation}
\int_{-a}^{+a} dx\; e^{itx} = 2\frac{\sin at}{t},
\end{equation}
such that
\begin{equation}
\begin{split}
& \Delta^{\rho\sigma}(x,y)= 4\int \frac{dq_3}{2\pi} \int \frac{ds_3}{2\pi}
e^{i(s_3y_3-q_3x_3)} L(a,s_3,q_3)
\int \frac{d^3\hat q}{(2\pi)^3}\; e^{i\hat q(\hat y -\hat x)}
\Delta^{\rho\mu}(\hat q,q_3) \Delta^{\sigma\nu}(\hat q, s_3)\;
i\Pi_{\mu\nu}(\hat q,B),\cr
&\hskip 3cm with\ L(a,s_3,q_3)=\int_{-\infty}^{+\infty} \frac{dp_3}{2\pi}  \frac{dr_3}{2\pi}
\;\frac{\sin a(q_3+p_3-r_3)}{q_3+p_3-r_3}
\; \frac{\sin a(r_3-p_3-s_3)}{r_3-p_3-s_3}.
\end{split}
\end{equation}

Going from the variables $r_3,p_3$ to the variables $p_3, h_3=r_3-p_3$ one
gets
\begin{equation}
L(a,s_3,q_3)= \int_{-\infty}^{+\infty} \frac{dp_3}{2\pi}\; K(a,s_3,q_3),\quad
with\ K(a,s_3,q_3)= 
\int_{-\infty}^{+\infty} \frac{dh_3}{2\pi}\;\frac{\sin
a(q_3-h_3)}{q_3-h_3}\; \frac{\sin a(h_3-s_3)}{h_3-s_3},
\end{equation}
and the photon propagator at 1-loop writes
\begin{equation}
\begin{split}
\hskip -1cm
\Delta^{\rho\sigma}(a,x,y) &= 4
\int_{-\infty}^{+\infty} \frac{d^3\hat q}{(2\pi)^3}\,e^{i\hat q(\hat y-\hat
x)}
\int_{-\infty}^{+\infty} \frac{ds_3}{2\pi}
\int_{-\infty}^{+\infty} \frac{dq_3}{2\pi} \;
e^{i(s_3y_3-q_3x_3)}\,
\Delta^{\rho\mu}(\hat q, q_3)\;K(a,s_3,q_3)\;
\Delta^{\nu\sigma}(\hat q, s_3) \;[\mu]\,
\Pi_{\mu\nu}(\hat q,B),\cr
& \hskip 2cm with\ \mu = \int_{-\infty}^{+\infty} \frac{dp_3}{2\pi}.
\end{split}
\label{eq:prop1}
\end{equation}
Last, going to the variable $k_3=s_3-q_3$
(difference of the 3-momentum of incoming and outgoing photon) , one gets
\begin{equation}
K(a,s_3,q_3)\equiv \tilde K(a,k_3)
 = \frac12 \frac{\sin a(s_3-q_3)}{s_3-q_3}
= \frac12 \frac{\sin ak_3}{k_3}.
\label{eq:Kexp}
\end{equation}
After truncating the external photon propagators, one can therefore define an
``effective vacuum polarization''
\begin{equation}
\Pi_{\mu\nu}^{eff}(q)= 4\mu\int_{-\infty}^{+\infty} \frac{dk_3}{2\pi}  \;
e^{ik_3 y_3}\, \tilde K(a,k_3)\;
(\Delta^{\nu\sigma}(\hat q, q_3))^{-1}\Delta^{\nu\sigma}(\hat q, q_3+k_3)
 \; \Pi_{\mu\nu}(\hat q,B),
\label{eq:Pieff}
\end{equation}
the meaning of $(\Delta^{\nu\sigma}(\hat q, q_3))^{-1}$ being that
$\Delta^{\nu\sigma}(\hat q, q_3)(\Delta^{\nu\sigma}(\hat q,
q_3))^{-1}\Delta^{\nu\sigma}(\hat q, q_3+k_3)= \Delta^{\nu\sigma}(\hat q,
q_3+k_3)$.

Since we have ``localized'' electrons inside graphene, we shall
conservatively consider
\begin{equation}
p_3\in[-\frac{(\hbar)}{a},+\frac{(\hbar)}{a}]
\quad\stackrel{def}{\Leftrightarrow}\quad
 p_3^m=\frac{(\hbar)}{a},
\label{eq:p3mdef}
\end{equation}
which amounts to take
\begin{equation}
\mu \approx \frac{1}{2\pi}\;\frac{2(\hbar)}{a} = \frac{(\hbar)}{a\pi}.
\label{eq:rhoval}
\end{equation}
We work in a system of units where $\hbar=1$ such that
\begin{equation}
\begin{split}
\Pi_{\mu\nu}^{eff}(q) &= \frac{1}{\pi^2}\;\Pi_{\mu\nu}(\hat q,B)
\times  U(q,y_3),\cr
U(q,y_3) &= \int_{-\infty}^{+\infty} dk_3  \;
e^{ik_3 y_3}\, \frac{\sin ak_3}{ak_3}\;
(\Delta^{\nu\sigma}(\hat q, q_3))^{-1}\Delta^{\nu\sigma}(\hat q,
q_3+k_3),
\end{split}
\label{eq:Pieff2}
\end{equation}
in which we have used the property  that
 $\Pi_{\mu\nu}(\hat q,B)$ can be taken out of the integral.
This demonstrates the result that has been announced and introduces the
transmittance function $U(q,k_3)$ which is independent of $B$.

Notice that:\newline
*\ the 1-loop photon propagator (\ref{eq:prop1}) still depends on the
difference $\hat y -\hat x$ but no longer depends on $y_3-x_3$ only,
it is now a function of both $y_3$ and $x_3$ (as already mentioned at the
end of subsection \ref{subsec:lightcone}, this ``extra'' dependence is in practice
very weak);\newline
*\ the ``standard'' calculation corresponds to $\hat
K(x)=\delta(x) \Rightarrow L(a,s_3,q_3)= \int_{-\infty}^{+\infty}
\frac{dp_3}{2\pi}\,
\frac{dr_3}{2\pi}\; \delta(q_3+p_3-r_3)\delta(r_3-p_3-s_3)
=\int \frac{dp_3}{2\pi}\; \delta(q_3-s_3)$.
Now, instead, we do not have momentum conservation along $z$. In
particular, while $\hat q=\hat s$, we do not have the relation
$q_3=s_3$;\newline
*\ when $a\to\infty$, there is momentum conservation along $B$ while for
finite $a$ it is only approximate.

\subsubsection{The transmittance function $\boldsymbol{U(q,y_3)}$.
A choice of gauge.}
%----------------------------------------------------------------

To get our final expression for the transmittance $U$, we shall hereafter
work in the Feynman gauge for the photons in which their propagators are
\begin{equation}
\Delta^{\mu\nu}(q) = -i\;\frac{g^{\mu\nu}}{q^2}.
\end{equation}
 Then, $U$ can be taken as
\begin{equation}
U(q,y_3)=\int_{-\infty}^{+\infty} dk_3  \;
e^{ik_3 y_3}\, \frac{\sin ak_3}{ak_3}\;
\frac{q_0^2-q_1^2-q_2^2 -q_3^2}{q_0^2-q_1^2-q_2^2 -(q_3+k_3)^2},
\label{eq:Pieff4b}
\end{equation}
in which we recall that the integration variable is $k_3=s_3-q_3$, the
momentum difference along $B$ between the outgoing and incoming photons.

The analytical properties and pole
structure  of the integrand in the complex $k_3$ plane will be seen to
 play an essential
role, like for the transmittance in optics (or electronics). This is why,
in addition to its ``classical'' and ``geometric'' character,  we have
given the same name to $U$.

\subsubsection{Going to dimensionless variables}
%-----------------------------------------------

It is time to go to dimensionless variables. We define ($p_3^m$ is given in
(\ref{eq:p3mdef}))
\begin{equation}
\eta=aq_0=\frac{q_0}{p_3^m},\quad
\zeta= a\sqrt{2eB} =\frac{\sqrt{2eB}}{p_3^m},\quad
\Upsilon= \frac{\eta}{\zeta} \gg 1,\quad
u=\frac{y_3}{a}.
\end{equation}
It is also natural, in $U$, to go to the integration variable
$\sigma=\frac{k_3}{p_3^m}$, and to introduce the refractive index $n$ and
the angle of incidence $\theta$ according to
\begin{equation}
q_2=0,\quad
q_1= |\vec q| s_\theta = n q_0 s_\theta,\quad
q_3= |\vec q| c_\theta = n q_0 c_\theta,\quad \theta \in ]0,\frac{\pi}{2}[,
\end{equation}
which, going to the integration variable $\sigma = ak_3=\frac{k_3}{p_3^m}$,
 leads to
\begin{equation}
U(q,y_3)=\frac{1-n^2}{a} V(u,n,\theta,\eta),\quad
V(u,n,\theta,\eta)  =\int_{-\infty}^{+\infty}
d\sigma\; e^{i\sigma u}\;
\frac{\sin\sigma}{\sigma}\frac{1}{1-n^2-\frac{\sigma}{\eta}
(2n\cos\theta+\frac{\sigma}{\eta})},
\label{eq:UVdef}
\end{equation}
and, therefore, to
\begin{equation}
\Pi_{\mu\nu}^{eff}(q) = \frac{1}{\pi^2}\;\Pi_{\mu\nu}(\hat q,B)
\frac{1-n^2}{a}\times  V(u,n,\theta,\eta).
\label{eq:PiV}
\end{equation}
We shall also call $V$ the transmittance function.

%SSSSSSSSSSSSSSSSSSSSSSSSSSSSSSSSSSSSSSSSSSSSSSSSSSSSSSSSSSSSSSSSSSSSSSSSSS
\section{The light-cone equations and their solutions} \label{section:lceqs}
%SSSSSSSSSSSSSSSSSSSSSSSSSSSSSSSSSSSSSSSSSSSSSSSSSSSSSSSSSSSSSSSSSSSSSSSSSS

\subsection{Orders of magnitude}\label{subsec:magnitude}
%=======================================================

In order to determine inside which domains we have to vary the dimensionless
parameters, it is useful to know the orders of magnitude of the
physical parameters involved in the study. 

$\bullet$\ The thickness of  graphene is $2a \approx 350\,pm = 350 \, 10^{-12}m$.

$\bullet$\  As we have seen in (\ref{eq:p3mdef}),
 $|p_3^{max}| \simeq \frac{(\hbar)}{a}$.
This gives $(c) \,p_3^{max} \simeq 1.8\,10^{-16} SI$ or
%$1\,GeV = 10^9 \times 1.6\,10^{-19} \,SI = 1.6\,10^{-10}\,SI$
%such that
$(c)\,p_3^{max} \simeq 1.13\,10^{-6}\,GeV = 1130\,eV \approx
2.2\,10^{-3}\,m_e$.

$\bullet$\  To $eB$ corresponds  $m^2 = \frac{(\hbar) e B}{(c)^2}$.
For example to $e B^m$ (see below) corresponds the mass $\frac{\sqrt{(\hbar) e
B^m}}{(c)} \approx
2~10^{-33}\,kg \approx 2~10^{-3}\,m_e \ll m_e$.
%since $m_e \approx 9.1\,10^{-31}\,kg$.

$\bullet$\ $[B] = \frac{[p]^2}{[e] (\hbar)}$ such that, to $(p_3^m)^2$
corresponds
$B^m \simeq \frac{(\hbar)}{e a^2} \approx 21531\,T$.

$\bullet$\ One has $\zeta\equiv\frac{\sqrt{2e(\hbar)B}}{p_3^m}=\sqrt{2\frac{B}{B^m}}$.
Since $B = \frac{\zeta^2}{2} B_m$, to $\zeta$ corresponds the
mass $\sqrt{2}\,\zeta \,10^{-3}\,m_e$.

$\bullet$\ 
%The earth magnetic field $B_t\in[.25,.65]\,G \equiv [.25, .65] \, 10^{-4}\,T$.
$1\,G \leftrightarrow \zeta \approx 9.64\,10^{-5}, 100\,G \leftrightarrow
\zeta \approx 9.64\,10^{-4}, 1\,T \leftrightarrow
\zeta \approx 9.64\,10^{-3}, 100\,T \leftrightarrow
\zeta \approx 9.64\,10^{-2}$.
\begin{equation}
1\,T \leq B \leq 20\, T \Leftrightarrow 1/100 \leq \zeta \leq
\sqrt{20}/100.
\label{eq:Brange}
\end{equation}

$\bullet$\ The wavelength of visible light lies between $350\,nm$ and $700\,nm$, that is
between
$3.5\,10^5\,pm$ and $7\,10^5\,pm$. For example light at $500 \,nm$
corresponds to an energy $\frac{(h c)}{\lambda} \approx
3.872\,10^{-19}\,SI \approx 2.48\,eV \approx (c)p_3^m/200 \ll (c)\,p_3^{max}$.
Likewise, light at  $350\,nm$ corresponds to  $3.54\,eV =
\frac{6.29}{1000}\;(c)p_3^m$, and
 at 800\,nm to $1.55\,eV = \frac{2.75}{1000}\; (c)p_3^m$.

So, the energy
of visible light $\ll (c)p_3^{max}$ and the corresponding $\eta$ satisfies
\begin{equation}
\text{visible light} \leftrightarrow 2.75\,10^{-3} \leq \eta \leq
6.3\,10^{-3}.
\label{eq:etarange}
\end{equation}

%$\bullet$\ In the middle of the visible spectrum $\eta \approx 5/1000$. To
%satisfy $\zeta \gg \eta$ we shall start at $\zeta \geq 1/100$.

\subsection{The light-cone equations}
%====================================

It is now straightforward to give the expression of the light-cone relations
(\ref{eq:lcperp}) and (\ref{eq:lcpar}) in the case of graphene.
First we express the relevant components of the vacuum polarization
$\Pi^{11}, \Pi^{22}, \Pi^{33}$ with dimensionless variables
\begin{equation}
\begin{split}
\Pi^{11} &= 4\alpha\, e^{-(n_x^2+n_y^2)\frac{\eta^2}{\zeta^2}}\; \zeta
\eta^2\,p_3^m \;\frac{n_x^2-n_y^2}{\eta^2-4\zeta^2},\cr
\Pi^{22} &= -\hat\Pi^{11},\cr
\Pi^{33} &= -4\alpha\, e^{-(n_x^2+n_y^2)\frac{\eta^2}{\zeta^2}}\;
\zeta p_3^m\; \frac{\zeta^2 -2(n_x^2+n_y^2)\eta^2}{\eta^2-4\zeta^2},
\end{split}
\label{eq:summary2}
\end{equation}
in which $n_x = n s_\theta$ and, since $q_2=0$, $n_y=0$.
(\ref{eq:PiV}) leads to
\begin{equation}
\begin{split}
& \star\ for\ A^\mu_\perp : (1-n^2)\left[1 +\frac{p_3^m}{\pi^2}\;\frac{1}{q_0^2}\;
\Pi^{22}(\alpha,n,\theta,\eta,\zeta)\; V(u,n,\theta,\eta)
\right]=0,\cr\;
& \star\ for\ A^\mu_\parallel : (1-n^2)\left[
1+\frac{p_3^m}{\pi^2}\;\frac{1}{q_0^2}
\Big( c^2_\theta\, \Pi^{11}(\alpha,n,\theta,\eta,\zeta) + s^2_\theta\,
\Pi^{33}(\alpha,n,\theta,\eta,\zeta)\Big)\;
V(u,n,\theta,\eta)\right]=0,
\end{split}
\label{eq:lcgeneral}
\end{equation}
and, using (\ref{eq:summary2}), to
\begin{equation}
\begin{split}
& \star\ for\ A^\mu_\perp : (1-n^2)\left[
1-\frac{4\alpha}{\pi^2}\,s_\theta^2 n^2\,
e^{-(n s_\theta\,\frac{\eta}{\zeta})^2}\;
\frac{\zeta}{\eta^2-4\zeta^2}\; V(u,n,\theta,\eta)\right]
=0,\cr
& \star\ for\ A^\mu_\parallel :
(1-n^2)\left[
1 + \frac{\alpha}{\pi^2}s_\theta^2
\left(4 c_\theta^2 n^2 \, \frac{\zeta}{\eta^2-4\zeta^2}
+\frac{\zeta}{\eta^2}
\frac{8\eta^2 n^2 s_\theta^2-4\zeta^2}{\eta^2-4\zeta^2}
\right)
\,e^{-(n s_\theta\,\frac{\eta}{\zeta})^2}\,V(u,n,\theta,\eta)\right]
=0.
\end{split}
\label{eq:lc4}
\end{equation}
This defines the index $n=n(\alpha, u,\theta,\eta,\zeta)$.

\subsection{Calculating the transmittance $\boldsymbol V$}
%=========================================================

In order to solve the light cone equations (\ref{eq:lc4}),
 the first step is to compute
$V$, so as to get an algebraic equation for $n$.
$V$ as given by (\ref{eq:UVdef})
 is the Fourier transform of the function $x \mapsto
-\eta ^2 \frac{\sin x}{x(x-\sigma_1)(x-\sigma_2)}$ where 
\begin{equation}
\sigma_1 = -\eta \left(n c_\theta - \sqrt{1-n^2 s_\theta^2}\right),\quad
\sigma_2 = -\eta \left(n c_\theta + \sqrt{1-n^2 s_\theta^2}\right).
\label{eq:poles}
\end{equation}
The Fourier transform of such a product of a cardinal sine with a
rational function is well known.
The result involves Heavyside functions of the imaginary parts of the
poles $\sigma_1, \sigma_2$, noted $\Theta_i^+$ for $\Theta_i
(\Im(\sigma_i))$ and $\Theta_i^-$ for $\Theta_i (-\Im(\sigma_i))$.
\begin{equation}
\hskip -1cm V(u, n, \theta, \eta)= \frac{-\pi \eta^2}{\sigma_1 \sigma_2 (\sigma_1 -
\sigma_2)}
\left[ (\sigma_1-\sigma_2)
+ \sigma_2 \left(\Theta_1^- e^{-i \sigma_1 (1-u)}+\Theta_1^+
e^{+i \sigma_1 (1+u)}\right)
-\sigma_1 \left(\Theta_2^- e^{-i \sigma_2
(1-u)}+\Theta_2^+ e^{+i \sigma_2 (1+u)}\right) \right].
\label{eq:transmit}
\end{equation}
The poles $\sigma_1, \sigma_2$
 are seen to control the behavior of $V$, thus  of $n$,
which depends  on the signs of their imaginary parts.

The $\frac{\sin a k_3}{ak_3} \equiv \frac{\sin\sigma}{\sigma}$ occurring in
$V$ (see (\ref{eq:UVdef}))
 provides, by its fast decrease, a natural cutoff in $k_3$ for the
integral, $|k_3| \leq \frac{1}{a}=p_3^m$. So, the amount of momentum
non-conservation of the photon in the direction of $B$ gets bounded by the
inverse of the confinement scale of electrons inside the graphene strip. 

The Fourier transform makes the transition between the momentum space in
which the propagators of the photons are written, and the position
space in which the evolution of the photons is described by the light-cone
equations.

It needs to be well defined,
which requires in particular that the poles be complex.
They are so when $n \not\in {\mathbb R}$ or when $n s_\theta >1$, that is
when $n_x >1$.

It cannot be applied when the poles are real, because the
integral is no more defined.
Then, in particular when $\theta \to 0$, the integral we shall
define as a Cauchy integral, like we did when calculating $\Pi^{\mu\nu}$,
arguing in particular of the $+i\varepsilon$ which is understood in the
denominator of the outgoing photon propagator. Then, $V$ will be
calculated through contour integration in the complex plane.

This alternate method can also be used when the poles are complex. It is
comforting that the 2 methods give, at leading order in an expansion at
small $\eta$ and $n_2$ ($n_2$ is the imaginary part of the refraction
index) the same results. In particular, the
cutoff that is then needed to stabilize the integration on the large upper
1/2 circle turns out to be the same as the one that naturally arises in the
Fourier transform because of the $\frac{\sin \sigma}{\sigma}$ function.

\subsection{Solving the light-cone equations for
$\boldsymbol{A^\mu_\parallel}$ and  $\boldsymbol{n\in {\mathbb
R}> \frac{1}{\sin\theta}}$}
%==============================================================

That $n \in {\mathbb R}$ largely simplifies the equations.

\subsubsection{Calculation of $\boldsymbol V$}
%---------------------------------------------

Expanding $V$ at leading orders in $\eta$, one gets
\begin{equation}
\begin{split}
\Re(V) &= -\frac{\pi}{\sqrt{ n^2 s^2_\theta-1}}\;\eta +
 \frac12 \pi(1 + u^2) \eta^2 + {\cal O}(\eta^3),\cr
\Im(V) &= u\, n\,c_\theta\;\frac{\pi}{\sqrt{ n^2 s^2_\theta-1}}\;\eta^2
+ {\cal O}(\eta^3).
\end{split}
\label{eq:Vexp}
\end{equation}
The expansion for $\Im(V)$ in (\ref{eq:Vexp}) starts at ${\cal
O}(\eta^2)$ while that of  $\Re(V) = {\cal O}(\eta)$.

For $n\in {\mathbb R} > \frac{1}{s_\theta}$ the 2 poles $\sigma_1$ and $\sigma_2$
(\ref{eq:poles}) of $V$ become
\begin{equation}
\sigma_1=-\eta\left(n \cos\theta -i  \sqrt{n^2 s_\theta^2 -1}\right), \qquad
\sigma_2=-\eta\left(n \cos\theta +i  \sqrt{n^2 s_\theta^2 -1}\right);
\label{eq:realpoles}
\end{equation}
the first term in $\Re(V)$ coincides with $\pm 2i\pi\times$ the
residue at the pole $\sigma_1$ or $\sigma_2$ (the one that lies inside the
contour of integration) when one calculates $V$ as a contour integral (see
also (\ref{eq:resapp})).

\subsubsection{The imaginary parts of the light-cone equations}
%--------------------------------------------------------------

The imaginary parts of both light-cone equations (\ref{eq:lc4}) shrink, for
$n$ real, to
\begin{equation}
\Im (V) =0.
\end{equation}
It is only rigorously satisfied at $u=0$, but, 
(\ref{eq:Vexp}) and  numerical
calculations show that, for values of $\eta$ in the visible spectrum
$\eta \in [3/1000, 7/1000]$,
$\Im(V) \ll \Re (V) <1$ and that $\Im(V)\approx 0$ is always an excellent
approximation.

\subsubsection{There is no non-trivial solution
for $\boldsymbol{A^\mu_\perp}$}
%----------------------------------------------

Detailed numerical investigations show that no solution exists for the
transverse polarization but the trivial solution $n=1$. We shall therefore
from now onwards only be concerned with photons $A^\mu_\parallel$ with a
parallel polarization (see Fig.\ref{fig:setup}).

\subsubsection{The light-cone equation for
$\boldsymbol{A_{\parallel}^{\mu}}$ and its solution }
%----------------------------------------------------

Expanding  $V$ in powers of $\eta$ and neglecting $\Im(V)$ enables to
get, through standard manipulations,
 a simple analytical equation for the refraction index $n$. For $\Upsilon
\gg 1$ and $\eta < 7/1000$, the following accurate expression is
obtained by expanding (\ref{eq:lc4}) in powers of $\frac{1}{\Upsilon}$
\begin{equation}
(1-n^2)\left[
1-\frac{\alpha}{\pi} \Upsilon\;\frac{s^2_\theta}{\sqrt{n^2s^2_\theta -1}}
\left(1+\frac{-3n^2s^2_\theta -c^2_\theta +1/4}{\Upsilon^2}
\right)\right] =0,
\label{eq:nparfull}
\end{equation}
which leads consistently  to the non-trivial solution
\begin{equation}
 n^2 \simeq \frac{1}{s_\theta^2}\ \frac{ 1+ \left(\frac{\alpha
 \Upsilon s_\theta^2}{\pi}\right)^2 \left( 1+\frac{1}{2\Upsilon^2}
\right)
 }{1+2\left(\frac{\alpha s_\theta}{\pi}\right)^2 (3s_\theta^2
 +c_\theta^2)},\quad \Upsilon=\frac{\sqrt{2eB}}{q_0}.
 \label{eq:solpar}
 \end{equation}

\subsubsection{Graphical results and comments}
%---------------------------------------------

\begin{figure}[h]
\begin{center}
\includegraphics[width=6 cm, height=5 cm]{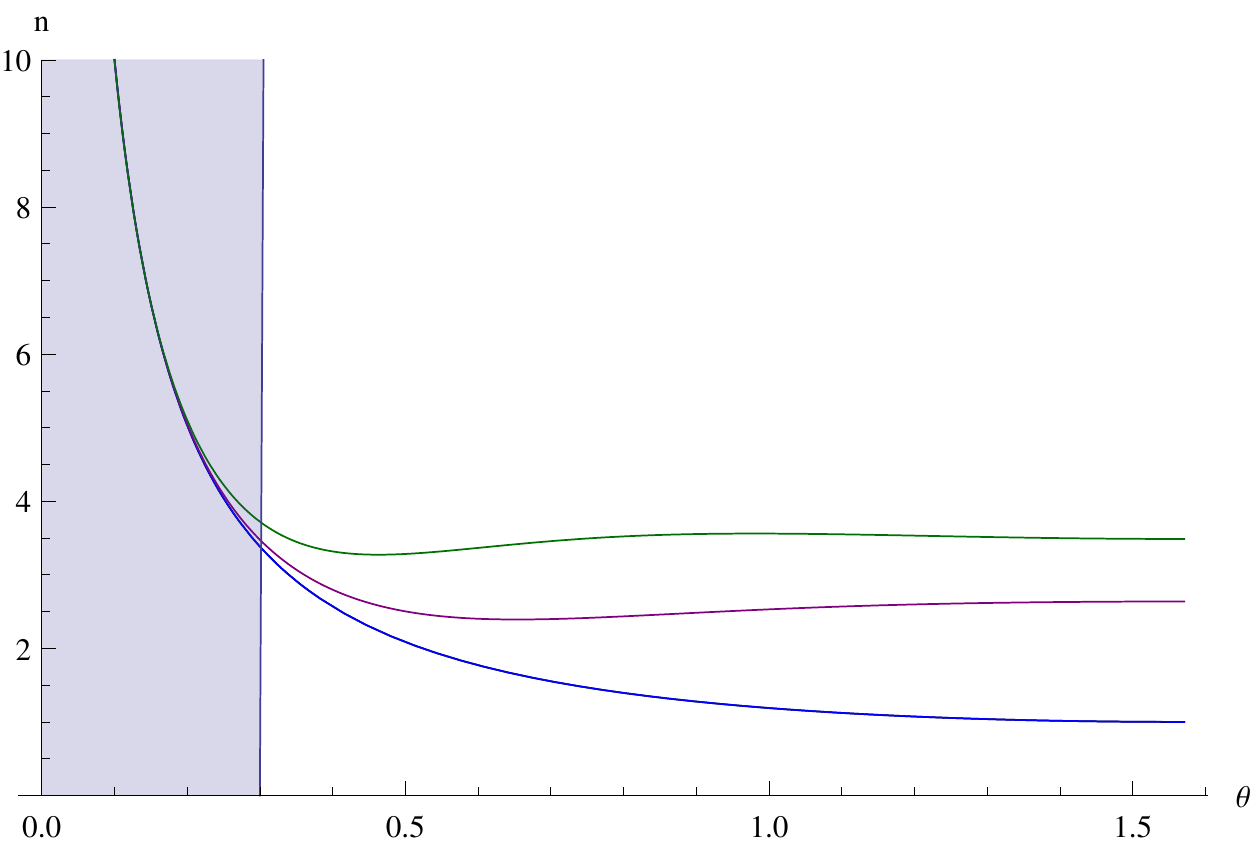}
\hskip 2cm
\includegraphics[width=6 cm, height=5 cm]{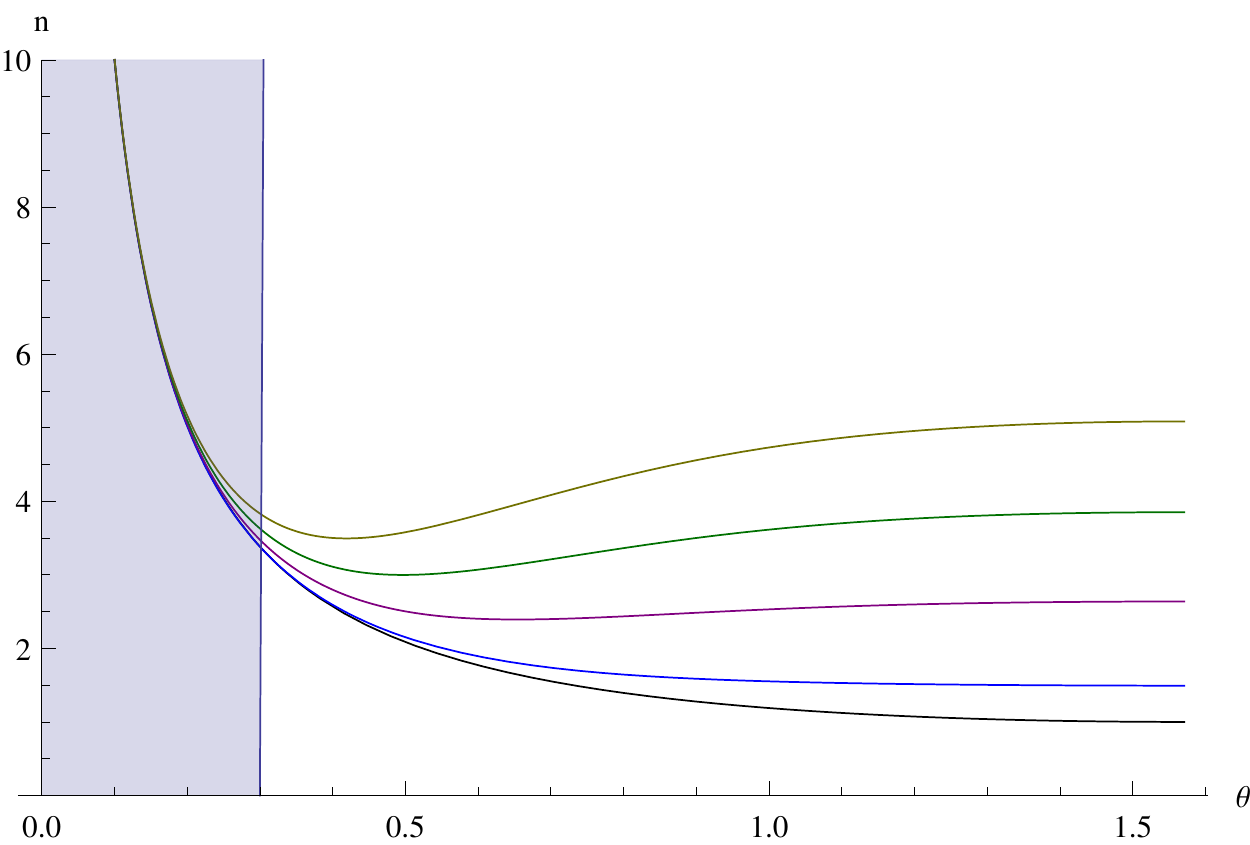}
\end{center}
\caption{The index $n\in{\mathbb R}$ for $A^\mu_\parallel$
 as a function of $\theta$. On the left we vary $\alpha =
1/137\,(blue), 1\,(purple), 2\,(green)$ at $\Upsilon=10$; on the right we vary
$\Upsilon = 5\,(blue), 10\,(purple), 15\,(green), 20\, (yellow)$ at $\alpha=1$. The
lower (black) curves are $1/\sin\theta$}
\label{fig:nBreal}
\end{figure}

The curves given by our final formula (\ref{eq:solpar}) are plotted on
Fig.\ref{fig:nBreal}.
On the left we vary $\alpha$ from $\frac{1}{137}$ to $2$
 at $\Upsilon = \frac{\sqrt{2eB}}{q_0}=10$ and on the right we keep
$\alpha=1$ and vary $\Upsilon$ between $5$ and $20$. On both plots, the
black lower curve in $n=\frac{1}{\sin\theta}$. We have shaded the domain of
low $\theta$ in which  $n$ must make a transition 
to another regime (see subsection \ref{subsec:limtetanull}).

$\bullet$\ The curves go asymptotically to $\frac{1}{s_\theta}$ when
$\theta \to 0$. However, we shall see that they should be truncated before
$\theta=0$).

$\bullet$\ At large angles, the effects are mainly of quantum nature, 
strongly influenced by the presence  of $B$ and largely depending on the
value of $\alpha$;
when $\theta$ gets smaller, one goes to
another regime in which the effects of confinement are the dominant ones.
The strict limit $\theta \to 0$ is special (see subsection
\ref{subsec:limtetanull}).

Quantum 1-loop effects are therefore potentially large at $\alpha
\geq 1$.  Furthermore, at reasonable values of $B$ and for photons
in the visible spectrum, the dependence on $B$ turns out to be strong. 
The ``confinement'' of  massless Dirac electrons inside a very
thin strip of graphene obviously acts as  an amplifier of the
effects of their interaction with  photons in a magnetic
background.

$\bullet$\ Quantum effects vary inversely  to the energy of
the photon : low frequencies are favored for testing, and this limit is
fortunate since our expansions were done precisely at $\eta = aq_0 \ll 1$.

$\bullet$\ For $\eta \ll 1$ and $n>\frac{1}{s_\theta}$,
 the residues of $V$ at the poles $\sigma_1$ and $\sigma_2$ are 
\begin{equation}
res(\sigma_1)=-\frac{\eta}{2i\sqrt{n^2 s^2_\theta-1}} + {\cal O}(\eta^2)
 = -res(\sigma_2).
\label{eq:resapp}
\end{equation}
The agreement between $\Re(V)$ in the first line of (\ref{eq:Vexp}) and $\pm
2i\pi\ res(\sigma_1)$ is conspicuous. Indeed, it is easy to prove that  for $n\in
{\mathbb R}$, only one of the 2 poles lies inside the contour of
integration in the upper 1/2 complex $\sigma$-plane
 which is the alternate method to calculate $V$.

This confirms that the transmittance function $U$ alone,
through its pole(s) is at the origin of the
``leading'' $\frac{1}{s_\theta}$ behavior of the refraction index (see
subsection \ref{subsub:leading}).
The poles are nothing more that the ones of the outgoing photon propagator
(in the Feynman gauge) $-i\frac{g^{\nu\sigma}}{\hat q^2 -(q^3+k_3)^2}$.
We recall that $k_3$ is the momentum non-conservation along $B$, which is
related (bounded by) to the momentum  allowed by quantum
mechanics to electrons confined into a strip of thickness $2a$. The
non-trivial poles of $U$ (or $V$) and the leading behavior of the
refraction index therefore originate  from the sole interactions
of  photons with ``confined''  electrons.

$\bullet$\ In the approximation that we made, the refractive index does not
depends on $u$, the position inside the strip. This dependence, very weak,
only starts to appear through higher orders in the expansion of the
transmittance $U$ (or $V$).

\subsubsection{The ``leading''
$\boldsymbol{n \sim \frac{1}{\sin\theta}}$ behavior}
\label{subsub:leading}
%---------------------------------------------------

It is easy to track the origin of the leading $\frac{1}{s_\theta}$
behavior of the index (we shall see below that the associated divergence at
$\theta \to 0$ is fake).

It comes in the regime when the 2 poles of $V$ lie in different 1/2 planes,
such that $V$ can be safely approximated by $V \approx 2i\pi\;
residue(\sigma_1\ or\ \sigma_2)$.

Keeping only the leading terms in the light-cone equation (\ref{eq:lc4})
and using (\ref{eq:resapp}) gives then
\begin{equation}
1 - \frac{\alpha}{\pi^2} \frac{\zeta}{\eta^2} \left(2i\pi 
\frac{\eta}{2i\sqrt{n^2s^2_\theta -1}}\right)=0,
\end{equation}
which yields
\begin{equation}
n^2s^2_\theta - 1 \sim \left(\frac{\alpha s^2_\theta \Upsilon}{\pi} \right)^2.
\end{equation}

The $\frac{1}{s_\theta}$ leading behavior of the index is therefore
associated with the transmittance $V$ and is  of
``geometric'' origin (shape of the sample, localization of the interaction
vertices inside the graphene strip). This gets confirmed in section
\ref{section:noB} where a similar study is done in the absence of any
external $B$: only the $\frac{1}{s_\theta}$ behavior of the index is then
practically left over.

\subsection{The transition $\boldsymbol{\theta \to 0}$}\label{subsub:transit}
\label{subsec:limtetanull}
%===========================================================================

It is fairly easy to determine the value of $\theta$ below which our
calculations and the resulting approximate formula (\ref{eq:solpar}) may
not be trusted anymore.
There presumably starts a transition to another regime.

Our calculations stay valid as long as the 2 poles $\sigma_1$ and
$\sigma_2$ of the transmittance function $V$ lie in  different 1/2 planes.
This requires that their imaginary parts have opposite signs. Their
explicit expressions are given in (\ref{eq:impoles}) below. It is then
straightforward to get the following condition
\begin{equation}
\sigma_1\ \text{and}\ \sigma_2\ \text{in  different 1/2 planes} \Leftrightarrow
n_1^2 > \frac{1+n_2^2}{\tan^2\theta}.
\label{eq:condval}
\end{equation}
(\ref{eq:condval}) is always satisfied at $\theta=\frac{\pi}{2}$ and never
at $\theta=0$. Since $n_2 \approx 0$, the transition occurs at 
\begin{equation}
n_1(\theta) \approx n(\theta) \approx \frac{1}{\tan\theta},
\end{equation}
in which we can use (\ref{eq:solpar}) for $n$. Since at small $\theta$,
$\sin\theta \simeq \theta \simeq \tan\theta$, this condition writes
approximately
\begin{equation}
1\leq  \frac{1+\left(\frac{\alpha\Upsilon s_\theta^2
}{\pi}\right)^2\left(1+\frac{1}{2\Upsilon^2}\right)}
{1+2\left( \frac{\alpha s_\theta}{\pi}\right)^2 (3s^2_\theta+c^2_\theta)}
\Leftrightarrow \theta \geq \theta_{min}= \sqrt{\frac{2}{\Upsilon^2-\frac72}}.
\label{eq:thetalim}
\end{equation}
For example, at $\Upsilon =5$ it yields $\theta \geq .3$.
Notice that the
condition (\ref{eq:thetalim}) also sets a lower limit $\Upsilon >
\sqrt{\frac72}$. 

It is easy to get the value  $n_{max}$ of $n$ at $\theta = \theta_{min}
\simeq \frac{\sqrt{2}}{\Upsilon}$ given by (\ref{eq:thetalim}). Plugging
this value in (\ref{eq:solpar}) one gets
\begin{equation}
n_{max} \equiv n(\theta =\theta_{min})\approx \frac{\Upsilon}{\sqrt{2}}.
\label{eq:nmax}
\end{equation}

Seemingly, the solution (\ref{eq:solpar})
 that we have exhibited gets closer and closer to
the ``leading'' $\frac{1}{s_\theta}$ when $\theta$ becomes smaller and
smaller.
The easiest way to show that this divergent is fake relies on a physical
argument: the poles of the outgoing photon propagator, which are also those
of the transmittance $U$ should be such that $|k_3|$, the momentum exchanged
with electrons along $B$ is smaller or equal than $\frac{1}{a} = p_3^m$,
which is the maximum quantum momentum of the confined electrons of
graphene. Mathematically, this traduces for the poles (\ref{eq:poles}) of $V$ by
\begin{equation}
|\sigma_1| \leq 1,\quad |\sigma_2| \leq 1.
\label{eq:phiscond}
\end{equation}
For $n > \frac{1}{s_\theta}$, both conditions yield
\footnote{
for $n < \frac{1}{s_\theta}$, the condition
$nc_\theta \leq \sqrt{1-n^2s^2_\theta}$ must also hold, and then one must have
$n^2 \leq 1$ (the case $nc_\theta \geq\sqrt{1-n^2s^2_\theta}$ or,
equivalently $n^2 \geq 1$ has no solution).} 
\begin{equation}
n^2 \leq n^2_{quant} =\frac{1}{\eta^2} +1.
\label{eq:nquant}
\end{equation}

Remark that $n_{max}$ is much smaller than the quantum limit
(\ref{eq:nquant}).

The case $\theta=0$ is special and is investigated directly. One has then
$\sigma_1 = -\eta(n-1), \sigma_2 = -\eta(n+1)$, such that $|\sigma_1|,
|\sigma_2| \leq 1$, that is
\begin{equation}
n(\theta=0) \stackrel{quantum}{\leq} \frac{1}{\eta}-1.
\label{eq:boundtetanul}
\end{equation}
For finite
$\eta$, this bound does not diverge, which shows that the diverging solution
(\ref{eq:solpar}) cannot be relied on down to $\theta =0$.
It can be trusted at most down to a value of $\theta$ for which $n^2
=n^2_{quant}$.
Therefore,  if a solution exists at $\theta=0$, $n$ must
cross the curve $n=\frac{1}{s_\theta}$ somewhere at small $\theta$.

However, as we now argue, such a transition cannot exist.
This is most easily proved by showing that, at no value of
$\theta$,
$n=\frac{1}{s_\theta}$ can be a solution to the light-cone equation
(\ref{eq:lc4}).
Let us write $\sigma_1 = -\eta
\frac{c_\theta}{s_\theta}+\epsilon, \sigma_2 =
-\eta\frac{c_\theta}{s_\theta}-\epsilon$. The poles being real, $V$ can be
calculated by setting  $\Theta(0)=\frac12$ in (\ref{eq:transmit}), which
yields
\begin{equation}
\begin{split}
V & \stackrel{real\ poles}{\rightarrow}
-\frac{\pi \eta^2}{\sigma_1\sigma_2(\sigma_1-\sigma_2)}
\left(\sigma_1-\sigma_2 +\sigma_2 \cos\sigma_1 e^{i\sigma_1 u}-\sigma_1
\cos\sigma_2 e^{i\sigma_2 u}\right)\cr
&\hskip -1cm =-\frac{\pi \eta^2}{\sigma_1\sigma_2(\sigma_1-\sigma_2)}
\Big( \sigma_1-\sigma_2
+ \sigma_2 \cos \sigma_1 \cos u\sigma_1 -\sigma_1 \cos\sigma_2 \cos
u\sigma_2
+i\big(\sigma_2 \cos\sigma_1 \sin u\sigma_1 -\sigma_1 \cos\sigma_2 \sin
u\sigma_2\big)
\Big).
\end{split}
\label{eq:Vlim}
\end{equation}
and, in our case, at $u=0$,
\begin{equation}
V(u=0) \approx -\pi \frac{s^2_\theta}{c^2_\theta}\left( 1-\cos(\eta^2
\frac{c^2_\theta}{s^2_\theta})-\eta\frac{c_\theta}{s_\theta}
\sin(\eta\frac{c_\theta}{s_\theta})\right).
\end{equation}
The light-cone equation (\ref{eq:lc4}) for $A^\mu_\parallel$ writes then
\begin{equation}
\left(1-\frac{1}{s^2_\theta}\right)\left[
1+\frac{\alpha}{\pi}\frac{s^2_\theta}{c^2_\theta}\frac{1}{\zeta} \left(
c^2_\theta -\Upsilon^2 s^2_\theta (1-\frac{2}{\Upsilon^2})\right)
\left(1-\cos(\eta^2\frac{c^2_\theta}{s^2_\theta})
-\eta\frac{c_\theta}{s_\theta}\sin(\eta\frac{c_\theta}{s_\theta})\right)
\right]=0,
\label{eq:lclim}
\end{equation}
in which we have incorporated the ``trivial'' term $(1-n^2)$.

 Eq.~(\ref{eq:lclim}) has no solution:
the crossing that would make the connection between our diverging solution
and an hypothetical solution in  the domain lying below the absolute
quantum bound (\ref{eq:nquant}) cannot be realized
\footnote{We have even investigated the existence of such solutions using
the exact expression for $V$, with the same conclusion. One has to be
careful that, in this case, the 2 poles are equal, and the expression of
$V$ must therefore be adapted.}. Hence, the domain in
which we can trust our solution (\ref{eq:solpar}) cannot be extended down to
$\theta=0$
\footnote{Actually, we have extended our numerical calculations to values
of $\theta$ for which the 2 poles of $V$ lie in
the same 1/2 plane. They show that, in practice, the solution
(\ref{eq:solpar}) stays valid even in a small domain below $\theta_{min}$.}.

Does graphene become ``opaque'' to photons (total reflection)
 at very small $\theta$,  or is
this the sign that, for  more and more energetic photons and
larger and larger external magnetic fields, the simple model that we made
for graphene is no longer valid? We cannot decide in the framework of this
limited study.
``Something may happen'' to photons
 below a certain angle of incidence, but we must also keep in mind that we
only used an expansion of the vacuum polarization dangerously truncated to
1-loop in a situation where $\alpha \simeq 2$.

This investigation will be continued in subsection \ref{subsub:wallcomp} for
$n\in{\mathbb C}$ (see also the concluding subsection
\ref{subsec:smallangle}).

\subsection{The quantum upper bound $\boldsymbol{n<n_{quant}}$.
The threshold at $\boldsymbol{B=B^m}$}
\label{subsec:nquant}
%----------------------------------------------------

We have seen in (\ref{eq:nquant})
 that Quantum Mechanics sets an upper bound $n_{quant}$ for the index.
 It is a large value for optical
frequencies but, when the energy of photons $q_0= \frac{\eta}{a}$
 increases, $n_{quant}$ decreases accordingly, its asymptotic
value being $1$ for infinitely energetic photons.

Our calculations being only valid at large
$\Upsilon\equiv\frac{\sqrt{2eB}}{q_0} \gg 1$,
harder and harder photons need  larger and larger values of $B$
(that probably cannot be realized on earth).
Then, $\theta_{min}$ given in (\ref{eq:thetalim}) also
 decreases, while $n_{max}\equiv n(\theta_{min})$ given by
(\ref{eq:nmax}) increases. A point can be reached at which $n_{max}$
becomes equal, then larger than $n_{quant}$.
$n_{max} \sim n_{quant}$ occurs at $\eta \simeq \frac{\sqrt{2}}{\Upsilon}
\Leftrightarrow \zeta \simeq \sqrt{2}$, independently of $\eta$.
It corresponds (see subsection \ref{subsec:magnitude}) to
$B\simeq B^m \approx 21531\,T$. 
$B^m$ appears therefore as the (very large)
 magnetic field at which the two upper bounds
$n_{max}$ and $n_{quant}$ coincide. Still increasing $B$ would result in
$n_{max}$ exceeding the quantum limit, which is impossible. So, new
phenomena are  expected for $B > B^m$, which lie beyond the scope
of this work.

\subsection{Reliability of the
 approximation $\boldsymbol{F(x) \approx\frac{1}{1-x}}$ for
the electron propagator} \label{subsec:lowen}
%==========================================================

The approximation (\ref{eq:workapp}) that we made for the expression of the electron
propagator inside graphene (see subsection \ref{subsub:Fappro}) is only
valid for low energy electrons with $p_0 \leq 1.2 \sqrt{2eB}$.
Using subsection \ref{subsec:magnitude} for the orders of magnitude, the lowest
external magnetic field $B=1\,T$ that we consider corresponds to
$\zeta \approx \frac{1}{100}$ and therefore to an energy  
 $\approx \sqrt{2}\,\frac{1}{100}\, 10^{-3}m_e \approx
 7\, eV$. Accordingly, our approximation is reliable for electrons with energy
$p_0 \leq 10\,eV$. This is  satisfied inside graphene.

Note that the visible light that we send through graphene has also energy
$\leq 3.5\,eV$.

\subsection{Going to $\boldsymbol{n\in {\mathbb C}}$}
\label{subsec:complex}
%=========================================================================

\subsubsection{The case of $\boldsymbol{A^\mu_\parallel }$}
%----------------------------------------------------------

\begin{figure}[h]
\begin{center}
\includegraphics[width=6 cm, height=4 cm]{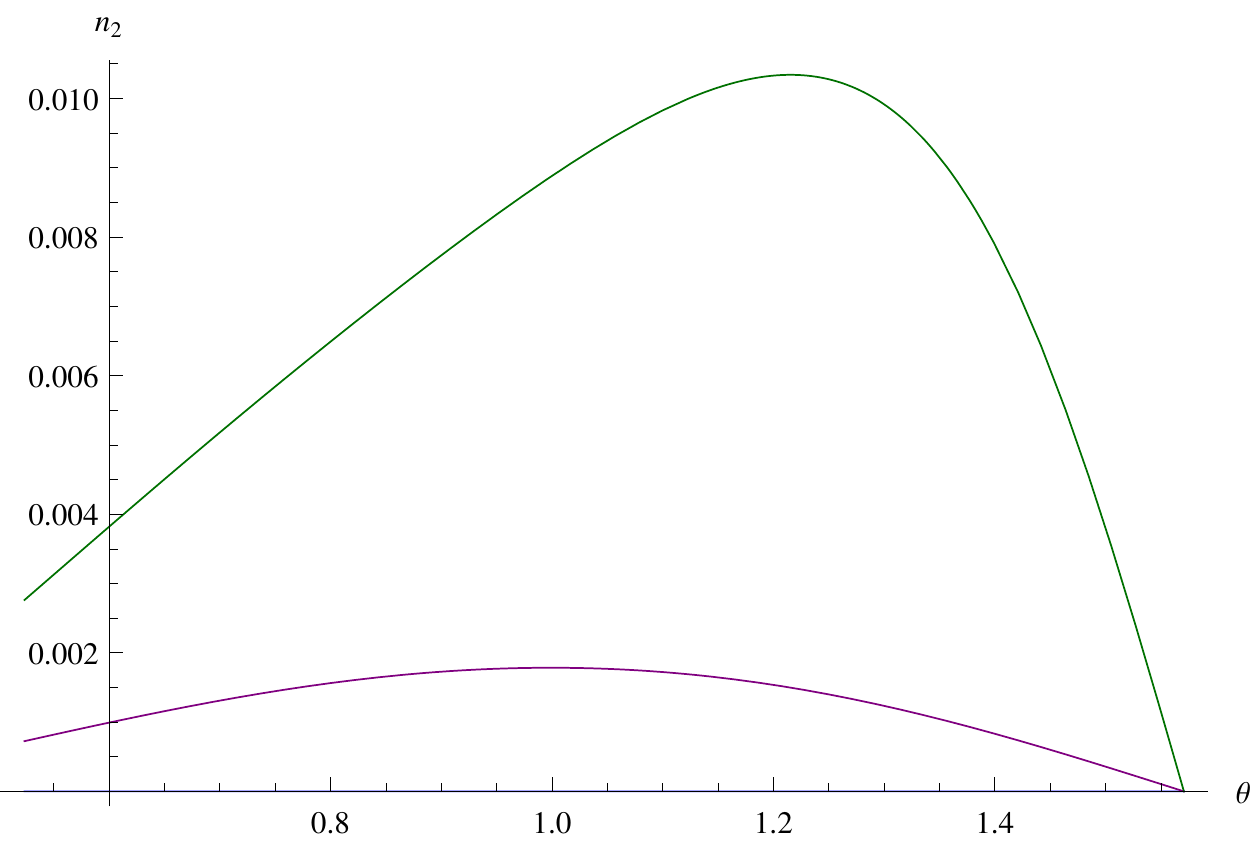}
\hskip 2cm
\includegraphics[width=6 cm, height=4 cm]{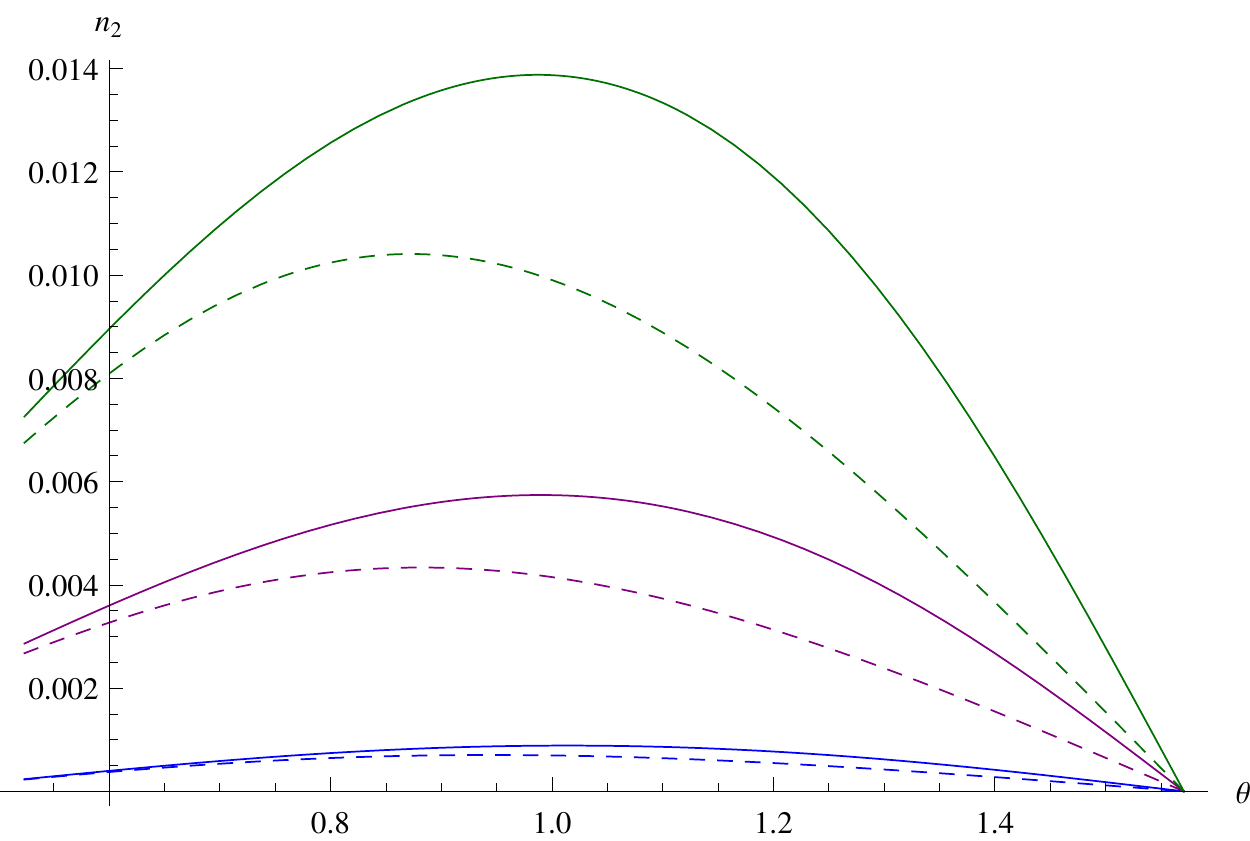}
\end{center}
\caption{The imaginary part $n_2$  of the  index $n$ for $A^\mu_\parallel$
 as a function of $\theta$. On the left
we vary $\alpha =
1/137\,(blue)$, $1\,(purple)$, $2\,(green)$ at $\Upsilon=5$; on the right we vary
$\Upsilon = 4\,(blue)$, $8\,(purple)$, $12\,(green)$ at $\alpha=1$. The dashed curves
on the right correspond to the rough approximation (\ref{eq:n2app})}
\label{fig:nBcomptheta}
\end{figure}

\begin{figure}[h]
\begin{center}
\includegraphics[width=6 cm, height=4 cm]{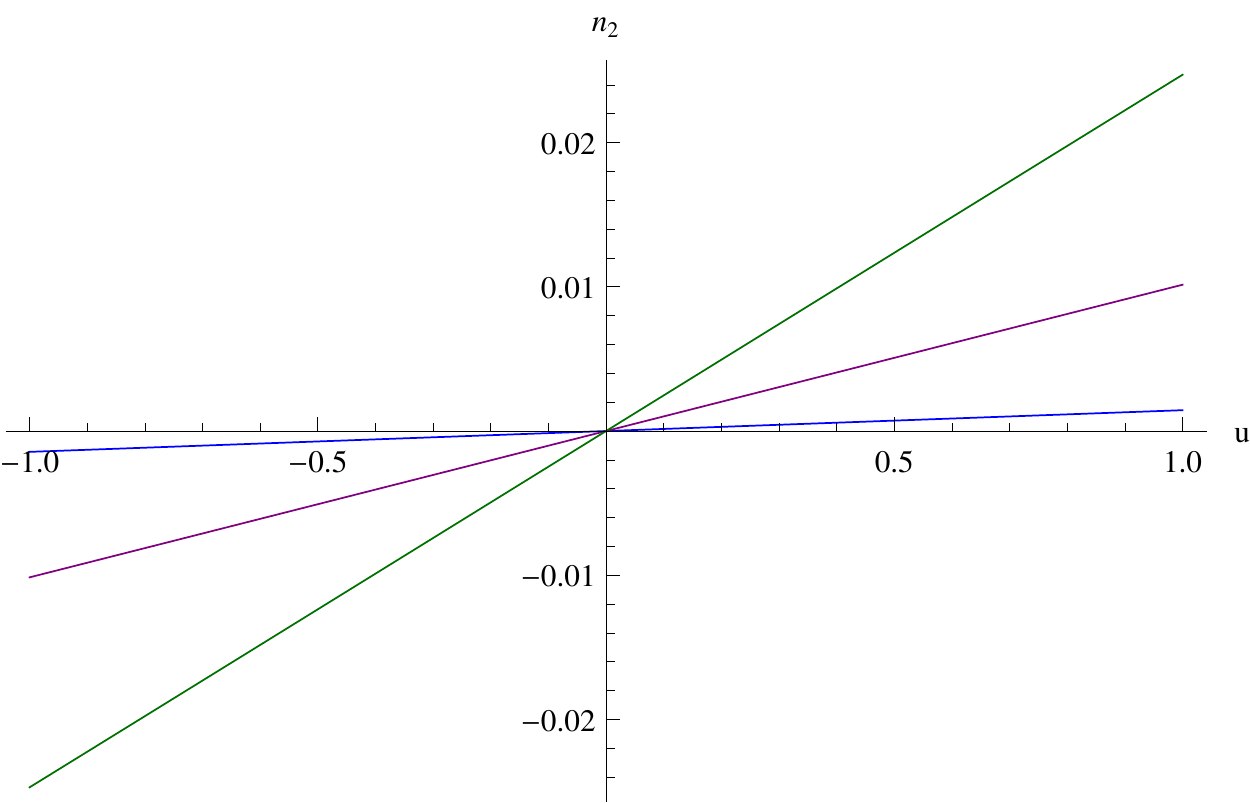}
\end{center}
\caption{The imaginary part $n_2$  of  index $n$ for $A^\mu_\parallel$
 as a function of $u$. We take $\alpha =1$, $\eta=5/1000$, and vary
$\Upsilon = 4\, (blue), 8\, (purple), 12\, (green)$}
\label{fig:nBcompu}
\end{figure}

Numerical calculations can be performed in the general case of a complex
index $n=n_1 + i n_2$. They show in particular that $|n_2| \ll n_1$,
confirming the reliability of the
 approximation that we made in the main stream of this study
(we have limited them to values of $\theta$ large enough for our equations
to be valid). The results are displayed on Fig.\ref{fig:nBcomptheta}, in
which we plot $n_2$ as a function of $\theta$, varying $\alpha$ (left) and
$\Upsilon$ (right), and on Fig.\ref{fig:nBcompu} in which we plot $n_2$ as
a function of $u$, varying $\Upsilon$.

To this purpose, and because the real part of the light-cone equation only
gets very slightly modified, it is enough to consider the imaginary part of the
light-cone equation (\ref{eq:lc4}) for $A^\mu_\parallel$
 in which we plug,  for $n_1^2$, the
analytic expression (\ref{eq:solpar}). In practice, the expansion of this
equation at ${\cal O}(\eta^2)$ and ${\cal O}(n_2)$, which is a polynomial
of first order in $n_2$ is enough for our purposes
An important ingredient of the calculation is the expansion of
the transmittance $V$ at order ${\cal O}(\eta^2)$ and ${\cal O}(n_2)$, in
the case when its 2 poles lie in different 1/2 planes,
which writes
\begin{equation}
\begin{split}
\frac{1}{\pi}\Re(V) &= -\frac{\eta}{\sqrt{n_1^2s^2_\theta
-1}}+\frac{1}{2}(1+u^2)\eta^2 + \frac{u c_\theta (2n_1^2s^2_\theta -1)
}{(n_1^2 s^2_\theta -1)^{\frac32}}\eta^2 n_2 + \ldots,\cr
\frac{1}{\pi}\Im(V) &= \frac{u n_1 c_\theta}{\sqrt{n_1^2 s^2_\theta -1}}\eta^2
-\frac{n_1 s^2_\theta}{(n_1^2 s^2_\theta -1)^{\frac32}}\eta n_2 + \ldots
\end{split}
\label{eq:Vexpcomp}
\end{equation}
The corresponding analytical expression for $n_2$, an odd function of $u$,
is long and  unaesthetic  and we only give it in  footnote \ref{foot:eqn2}
\footnote{The imaginary part of the light-cone equation for
$A^\mu_\parallel$ writes
\begin{equation}
\begin{split}
& M + N n_2 =0,\cr
& M =
u \zeta c_\theta s_\theta^2 (-1 + n_1^2 s_\theta^2) + 
 \frac{1}{4 \zeta}\eta^2 u c_\theta s_\theta^2
(1 - 4 n_1^2 c_\theta^2 - 12 n_1^2 s_\theta^2)
(-1 + n_1^2 s_\theta^2),\cr
& N =
-\frac{\zeta s_\theta^4}{\eta} - \frac{1}{\zeta}
  \eta^2 (1 + u^2) s_\theta^2 (c_\theta^2 + 3 s_\theta^2)
(-1 + n_1^2 s_\theta^2)^{\frac32} + \frac{1}{4 \zeta}
(-8 \eta c_\theta^2 s_\theta^2 - 25 \eta s_\theta^4 + 
    12 \eta n_1^2 c_\theta^2 s_\theta^4 + 36 \eta n_1^2 s_\theta^6).
\end{split}
\label{eq:eqn2}
\end{equation}
\label{foot:eqn2}}.
However a rough order of magnitude can be obtained with
very drastic approximations which lead to the equation
\begin{equation}
n_2 s^2_\theta \sim u \eta
c_\theta(n_1^2 s_\theta^2-1),
\label{eq:n2app}
\end{equation}
in which, like before, we can plug in the analytical formula
(\ref{eq:solpar}) for $n_1^2$. The corresponding curves  are the dashed ones in
Fig.\ref{fig:nBcomptheta}
\footnote{The agreement with the exact curves worsens as $\alpha$
increases.}.

As $B$ increases, it is no longer a reliable approximation to consider the
index to be real : absorption becomes non-negligible.
The window of medium-strong $B$'s from 1 to 20 Teslas together with photons
in the visible range appears therefore  quite simple and special.
Outside this window, the physics is most
probably much more involved and equations much harder to solve.

\subsubsection{The ``wall'' for $\boldsymbol{A^\mu_\parallel}$}
\label{subsub:wallcomp}
%--------------------------------------------------------------

The situation is best described in the complex $(n_1,n_2)$ plane of the
solutions $n=n_1+in_2$ of the light-cone equation (\ref{eq:lc4}) for
$A^\mu_\parallel$, which decomposes into its real and imaginary parts (in
the limit $\eta \ll \zeta \Leftrightarrow \Upsilon \gg 1$, and neglecting
the exponential $e^{-\frac{n^2 s^2_\theta}{\Upsilon^2}}$ which plays a
negligible role) according to
\begin{equation}
\begin{split}
& \ast\ 1+\frac{\alpha}{\pi}
\frac{s^2_\theta}{\zeta}\left(1+\frac{1}{4\Upsilon^2}\right)\left[\left(\Upsilon^2
-(n_1^2-n_2^2)(1+s^2_\theta)\right) \Re(V) +2n_1n_2(1+s^2_\theta)
\Im(V)\right]=0,\cr
& \ast\ -2n_1n_2 (1+s^2_\theta) \Re(V) +
\left(\Upsilon^2-(n_1^2-n_2^2)(1+s^2_\theta\right)\Im(V)=0.
\label{eq:lccomp}
\end{split}
\end{equation}

All previous calculations favoring solutions with
low absorption $|n_2| \ll n_1$, it is  in this regime that we shall
investigate the presence of a ``wall'' at small $\theta$. To this purpose,
we shall plug into the light-cone equation (\ref{eq:lc4}) for
$A^\mu_\parallel$ the expansion of the transmittance $V$ that is written in
(\ref{eq:Vexpcomp}).

\begin{figure}[h]
\begin{center}
\includegraphics[width=6 cm, height=5 cm]{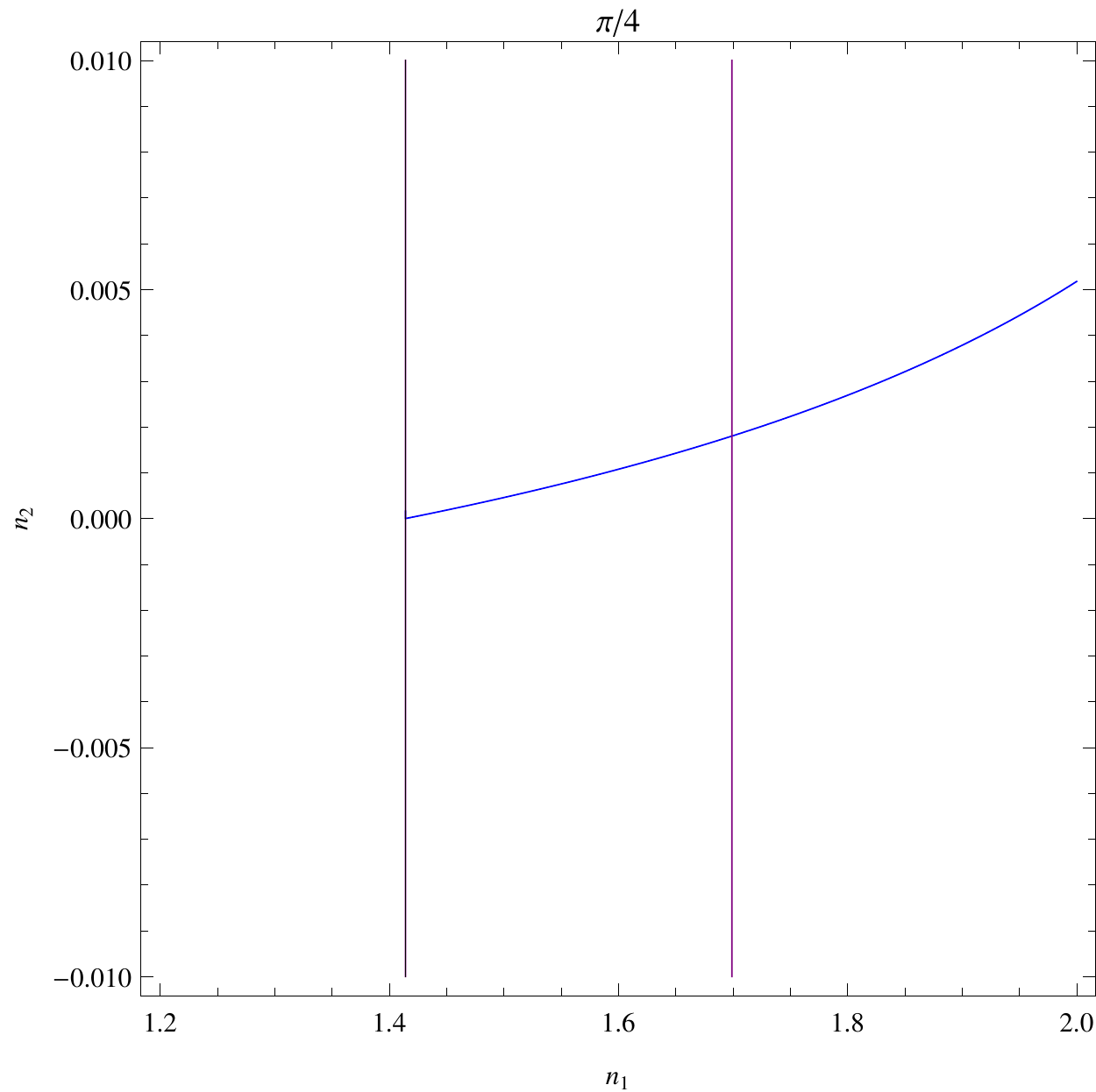}
\hskip 2cm
\includegraphics[width=6 cm, height=5 cm]{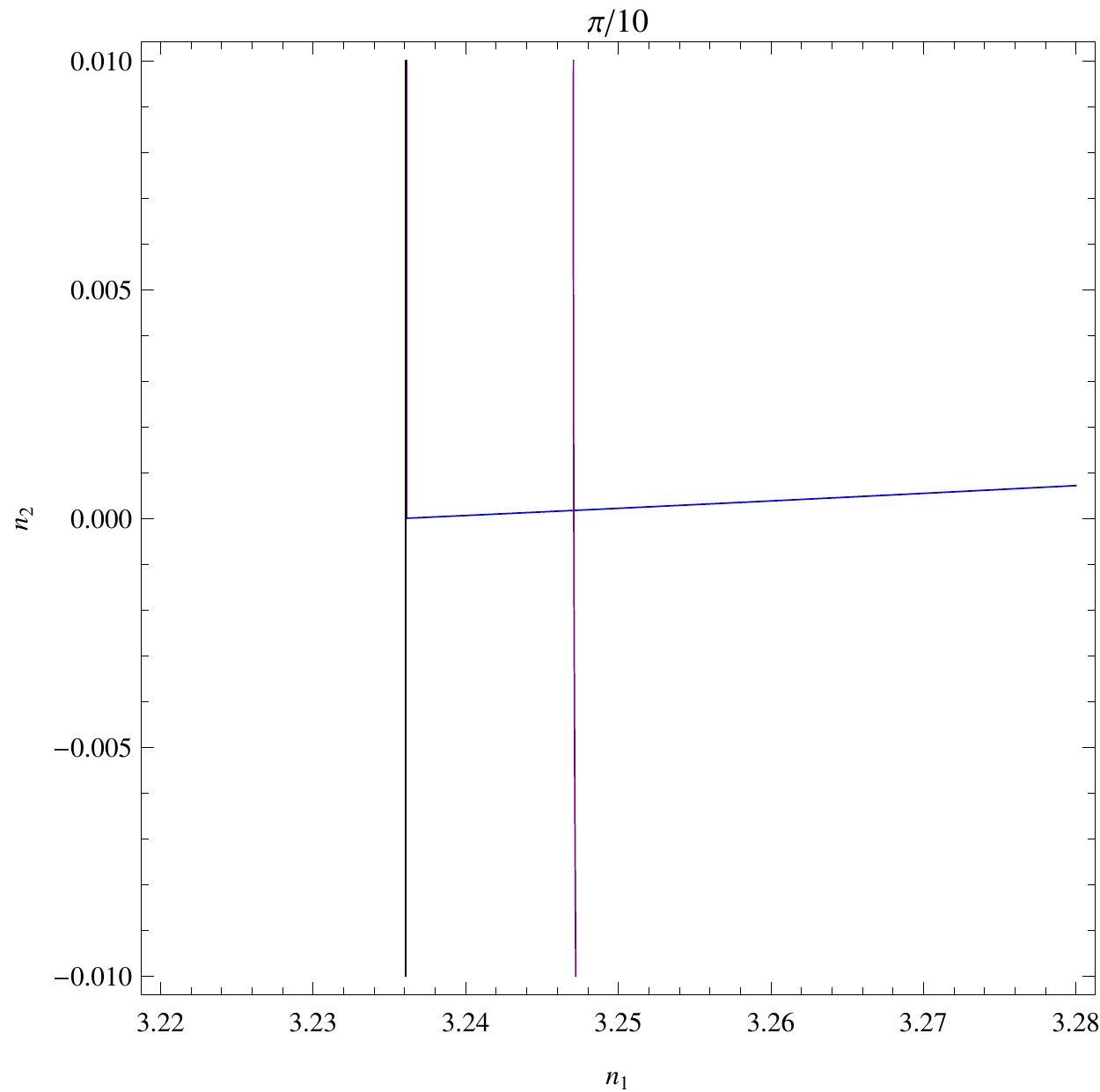}
\end{center}
\caption{The index $(n_1,n_2)$ for $A^\mu_\parallel$ at
$\theta=\frac{\pi}{4}$ (left) and $\theta=\frac{\pi}{10}$ (right)}
\label{fig:wall1}
\end{figure}

The situation at  $\theta=\frac{\pi}{4}$ (left) and $\theta=\frac{\pi}{10}$
 are depicted in Fig.\ref{fig:wall1}. The values
of the parameters are $\alpha=1, u=.5, \eta=\frac{5}{1000}, \Upsilon=5$.

The purple curve corresponds to the solutions of the real part of the
light-cone equation and the blue quasi-vertical line to the solution of its
real part. The intersection of the 2 curves yields the
solution $n=n_1+in_2$. We recover  $|n_2| \ll n_1$.
 The black vertical line on the left corresponds to $n_1=\frac{1}{s_\theta}$.

\begin{figure}[h]
\begin{center}
\includegraphics[width=6 cm, height=5 cm]{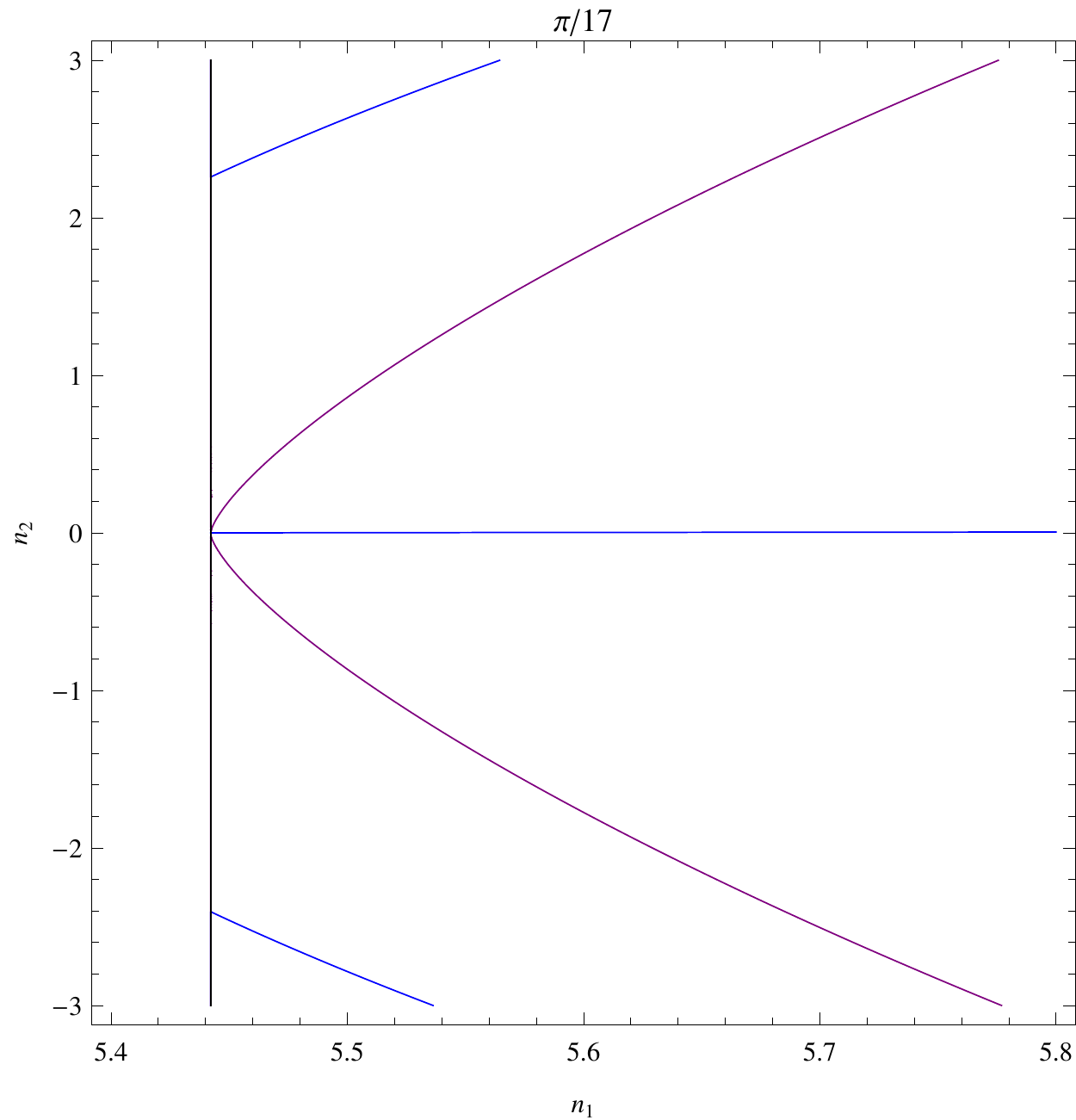}
\hskip 2cm
\includegraphics[width=6 cm, height=5 cm]{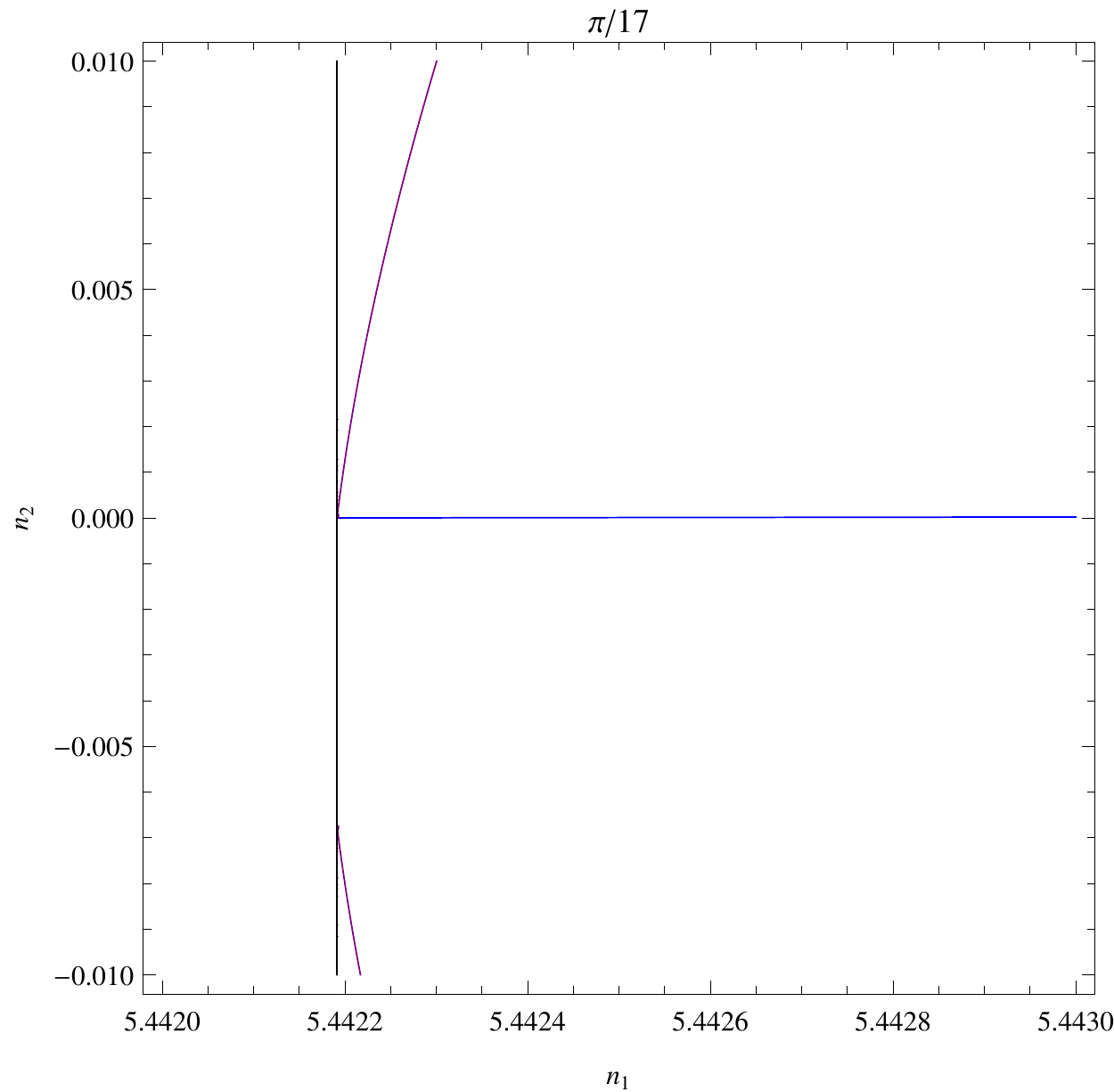}
\end{center}
\caption{The index $(n_1,n_2)$ for $A^\mu_\parallel$ at
$\theta=\frac{\pi}{17}$. The right figure is an enlargement of the left one}
\label{fig:wall2}
\end{figure}

A transition brutally occurs close to $\theta=\frac{\pi}{14}$. Then the
solution at $|n_2| \ll n_1 ={\cal O}(1)$ disappears. It is clearly visible on
Fig.\ref{fig:wall2} below in which we plot the situation after the transition, for
$\theta=\frac{\pi}{17}$.

There is no more intersection between the solutions of the real (purple)
and imaginary (blue) parts of the light-cone equations, except at $n_2=0,
n_1=\frac{1}{s_\theta}$, which is a fake solution since we know that $n_1$
can never reach its ``asymptotic'' value $\frac{1}{s_\theta}$.

\subsubsection{An estimate of the angle of transition
$\boldsymbol{\theta_{min}}$}
%----------------------------------------------------

This change of regime  is characterized by a brutal jump in the value of $n_2$,
which should be manifest on the imaginary part of the light-cone equation
(\ref{eq:lccomp}). A very reliable approximation can be obtained by
truncating $\Re(V)$ to its first term, in which case one gets
\begin{equation}
n_2 \approx (n_1^2 s^2_\theta-1)\frac{u\eta c_\theta\left(\Upsilon^2
-n_1^2(1+s^2_\theta)\right)}{s^2_\theta\left(\Upsilon^2
-n_1^2(1+s^2_\theta)\right)-2(1+s^2_\theta)(n_1^2 s^2_\theta-1)}
\end{equation}
which has a pole at (we use $s^2_\theta \ll 1$)
\begin{equation}
n_1^2 \approx \frac{2 + \Upsilon^2 s^2_\theta}{3s^2_\theta}.
\label{eq:n1max}
\end{equation}
This value for $n_1$ determines the maximum that can be reached when
$\theta$ decreases. Indeed, then, $n_2$ becomes out of control in the
framework of our approximations.
We also know that that $n_1$ should stay below $\frac{1}{s_\theta}$.
The intersection of (\ref{eq:n1max}) and
$\frac{1}{s_\theta}$ yields the lower limit for $\theta$
\begin{equation}
\theta_{min} \sim \frac{1}{\Upsilon}.
\label{eq:thetamin}
\end{equation}
(\ref{eq:thetamin}) is smaller than our previous estimate
(\ref{eq:thetalim}) obtained in the approximation $n\in{\mathbb R}$.

At $\Upsilon=5$ one gets $\theta_{min} \approx \frac{\pi}{15}$, which shows
the reliability of our estimate (the true transition numerically
 occurs between $\frac{\pi}{14}$ and $\frac{\pi}{15}$).

\subsubsection{The case of $\boldsymbol{A^\mu_\perp}$}
%-----------------------------------------------------

We only summarize below the steps that lead to the conclusion that no solution
to the refraction index except the trivial $n=1$ exists for the transverse
polarization.

Starting from the corresponding light-cone equation in (\ref{eq:lc4}), the
main task is to get the appropriate expression for the transmittance
function $V$.
To this purpose the starting point is the general expression
(\ref{eq:transmit}). We expand it in powers of $\eta$ in the sense that the
exponentials are expanded at ${\cal O}(\eta)$ or, eventually ${\cal
O}(\eta^2)$. No expansion in powers of $n_2$ is done because, if solutions
exist, they may occur for fairly larges values of $n_2$ (and $n_1$).

Since the sign of the imaginary parts of the poles $\sigma_1$ and
$\sigma_2$ obviously play a central role, it is also useful to extract
\begin{equation}
\begin{split}
\Im(\sigma_1) &= \eta \left(
-n_2 c_\theta + \frac{1}{\sqrt{2}}\sqrt{-c + \sqrt{c^2+d^2}}
\right),\cr
\Im(\sigma_2) &= \eta \left(
-n_2 c_\theta - \frac{1}{\sqrt{2}}\sqrt{-c + \sqrt{c^2+d^2}}
\right),\cr
 c &= 1-(n_1^2-n_2^2) s^2_\theta,\quad d= 2n_1n_2 s^2_\theta.
\end{split}
\label{eq:impoles}
\end{equation}

Straightforward manipulations on (\ref{eq:transmit}) show that:

* when $n_2 > 0$ ($\Rightarrow \Im(\sigma_2) <0$):
 if $\Im(\sigma_1)>0$, $V= \frac{-i\pi\eta}{\sqrt{1-n^2
s^2_\theta}}+\ldots$;
if $\Im(\sigma_1)<0$, $V= \frac{\pi\eta^2}{2}(1-u)^2+\ldots$\newline
* when $n_2 <0$  ($\Rightarrow \Im(\sigma_1) > 0$):
if $\Im(\sigma_2)>0$, $V=\frac{\pi \eta^2}{2} (1+u)^2+\ldots$;
if $\Im(\sigma_2) < 0$, $V =
\frac{-i\pi\eta}{\sqrt{1-n^2s^2_\theta}}+\ldots$

The cases when $V={\cal O}(\eta^2)$ correspond to $\sigma_1$ and $\sigma_2$
being in the same 1/2 complex $\sigma$-plane.

When $V=  \frac{-i\pi\eta}{\sqrt{1-n^2 s^2_\theta}}$,
its real and imaginary parts are given by
\begin{equation}
\Re(V) = \frac{\pi\eta n_1n_2 s^2_\theta}{\sqrt{2}}
\frac{\sqrt{c + \sqrt{c^2+d^2}}}{\sqrt{c^2 + d^2}},\qquad
\Im(V) = \frac{-\pi\eta}{\sqrt{2}} \frac{\sqrt{-c+\sqrt{c^2+d^2}}}
{\sqrt{c^2+d^2}}.
\end{equation}
Numerical solutions of the light-cone equation show that no solution exists
that fulfill the appropriate criteria on the signs of $\Im(\sigma_1),
\Im(\sigma_2)$. For example, for $n_2<0$, one gets  solutions shared by both
the real and imaginary parts of the light-cone equations, but they satisfy
$\Im(\sigma_2)>0$ and must therefore be rejected.

The next step is to use the exact expression (\ref{eq:transmit})  of $V$,
but no acceptable solution exists (solutions with very large values of
$n_1$ and $n_2$, larger than $20$, are a priori rejected).

%SSSSSSSSSSSSSSSSSSSSSSSSSSSSSSSSSSSSSSSSSSSSSSSSSSSSSSS
\section{The case $\boldsymbol{B=0}$}\label{section:noB}
%SSSSSSSSSSSSSSSSSSSSSSSSSSSSSSSSSSSSSSSSSSSSSSSSSSSSSSS

\subsection{The vacuum polarization $\boldsymbol{\Pi^{\mu\nu}}$}
%===============================================================

Standard techniques applied to massless electrons of graphene at the Dirac
point lead to the exact results
\begin{equation}
\begin{split}
i\Pi^{11}_{\xcancel B}(\hat q) &= i\frac{e^2}{8}\left(
\sqrt{\hat q_E^2}-\frac{q_1^2}{\sqrt{\hat q_E^2}}\right),\cr
i\Pi^{22}_{\xcancel B}(\hat q) &= i\frac{e^2}{8}\left(
\sqrt{\hat q_E^2}-\frac{q_2^2}{\sqrt{\hat q_E^2}}\right),\cr
i\Pi^{33}_{\xcancel B} &= i\frac{e^2}{4} \sqrt{\hat q_E^2},\cr
i\Pi^{00}_{\xcancel B} &= -i\frac{e^2}{8} \frac{q_1^2+q_2^2}{\sqrt{\hat
q_E^2}},\cr
i\Pi^{12}_{\xcancel B} &= -i\frac{e^2}{8} \frac{q_1q_2}{\sqrt{\hat
q_E^2}},\cr
i\Pi^{01}_{\xcancel B} &= -\frac{e^2}{8} \frac{q_0^E q_1}{\sqrt{\hat
q_E^2}},\cr
i\Pi^{02}_{\xcancel B} &= -\frac{e^2}{8} \frac{q_0^E q_2}{\sqrt{\hat
q_E^2}},\cr
& \Pi^{03}_{\xcancel B}=\Pi^{13}_{\xcancel
B}=\Pi^{23}_{\xcancel B}=0.
\end{split}
\label{eq:PinoB}
\end{equation}
in which $q_0=iq_0^E$ and $(\hat q^E)^2=(q_0^E)^2 + q_1^2 + q_2^2$.

$\star$\ $\Pi^{ij}_{\xcancel B}$ is proportional to $\pi \alpha$
while, in the presence of $B$, it was proportional to $\alpha$. The extra
$\pi$ comes from $\int_0^1 dx\;\sqrt{x(1-x)}= \frac{\pi}{8}$.

$\star$\ Transversality: one easily checks on (\ref{eq:PinoB}) that
$q_0\Pi^{00}_{\xcancel B}+q_1\Pi^{10}_{\xcancel B}+q_2\Pi^{20}_{\xcancel B}
= 0$, $q_0\Pi^{01}_{\xcancel B}+q_1\Pi^{11}_{\xcancel
B}+q_2\Pi^{21}_{\xcancel B}=0$,
$q_0\Pi^{02}_{\xcancel B}+q_1\Pi^{12}_{\xcancel B}+q_2\Pi^{22}_{\xcancel B}=0$. The last condition
$q_0\Pi^{03}_{\xcancel B} +q_1\Pi^{13}_{\xcancel B}+q_2\Pi^{23}_{\xcancel
B} +q_3\Pi^{33}_{\xcancel B}=0$
reduces to $q_3\Pi^{33}_{\xcancel B}=0$, which is not satisfied unless $q_3=0$ or
$\hat q_E^2=0$ (``on mass shell 2+1 photon'').

In our setup,
we recall $n^2=\frac{q_1^2 + \xcancel{q_2^2} + q_3^2}{q_0^2}$,
$n_x=\frac{q_1}{q_0}= ns_\theta, n_z=\frac{q_3}{q_0}= nc_\theta$.
One has $(\hat q_E)^2 = (q_0^E)^2 + q_1^2 + q_2^2=
q_0^2(n^2s^2_\theta-1)$ because $q_2=0$. So, $\sqrt{\hat q_E^2}
=  \pm q_0\sqrt{n^2s^2_\theta-1}$.  This gives
\begin{equation}
\begin{split}
\Pi^{11}_{\xcancel B} &= \mp \frac{\pi\,\alpha}{2} q_0
\frac{1}{\sqrt{n^2 s^2_\theta-1}},\cr
\Pi^{22}_{\xcancel B} &= \pm \frac{\pi\,\alpha}{2} q_0
\sqrt{n^2s^2_\theta-1},\cr
\Pi^{33}_{\xcancel B} &= \pm \pi\,\alpha\, q_0
\sqrt{n^2s^2_\theta-1},\cr
\Pi^{00}_{\xcancel B} &= \mp \frac{\pi\,\alpha}{2} q_0
\frac{n^2s^2_\theta}{\sqrt{n^2s^2_\theta-1}},\cr
\Pi^{12}_{\xcancel B} &=0,\quad \Pi^{i3}_{\xcancel B}=0,\cr
\Pi^{01}_{\xcancel B} &= \pm \frac{\pi \alpha}{2} q_1
\frac{1}{\sqrt{n^2 s^2_\theta-1}},\cr
\Pi^{02}_{\xcancel B} &= \pm \frac{\pi \alpha}{2} q_2
\frac{1}{\sqrt{n^2 s^2_\theta-1}}.
\end{split}
\label{eq:PinoB2}
\end{equation}

\subsection{The light-cone equation and the refractive index}
%============================================================

The light-cone equations (\ref{eq:lcgeneral}) together with (\ref{eq:PinoB2})
  yield
\begin{equation}
\begin{split}
& for\ A^\mu_\perp: (1-n^2)\left[ 1\pm
\frac{\alpha}{\eta}\sqrt{n^2s^2_\theta-1}
\;\frac{V(u,\rho,n,\theta,\eta)}{\pi}
\right]=0,\cr
&  for\ A^\mu_\parallel:(1-n^2)\left[1\pm \frac{\alpha}{\eta}
\Big(-\frac{c^2_\theta}{2\sqrt{n^2s^2_\theta-1}} +
s^2_\theta\sqrt{n^2s^2_\theta-1}\Big)
\;\frac{V(u,\rho,n,\theta,\eta)}{\pi}
\right]=0.
\end{split}
\label{eq:lcnoB}
\end{equation}
in which $V$ is the same transmittance function as before, given by
(\ref{eq:UVdef}).

\subsection{Solutions for $\boldsymbol{A^\mu_\parallel}$
 with $\boldsymbol{n\in{\mathbb R}}$}
%========================================================

We approximate, at $\eta \ll 1$, according to
(\ref{eq:Vexp}), $V \approx  -\frac{\eta\pi}{\sqrt{n^2s^2_\theta -1}}$.

Like in the presence of $B$, no non-trivial solution exists for the
transverse polarization and we focus hereafter on $A^\mu_\parallel$.
The corresponding light-cone equation writes
\begin{equation}
1\pm\alpha \left(-\frac{c_\theta^2}{2(n^2s_\theta^2-1)}+ s^2_\theta
\right)=0,
\label{eq:lcnoBpar}
\end{equation}

\begin{figure}[h]
\begin{center}
\includegraphics[width=6 cm, height=5 cm]{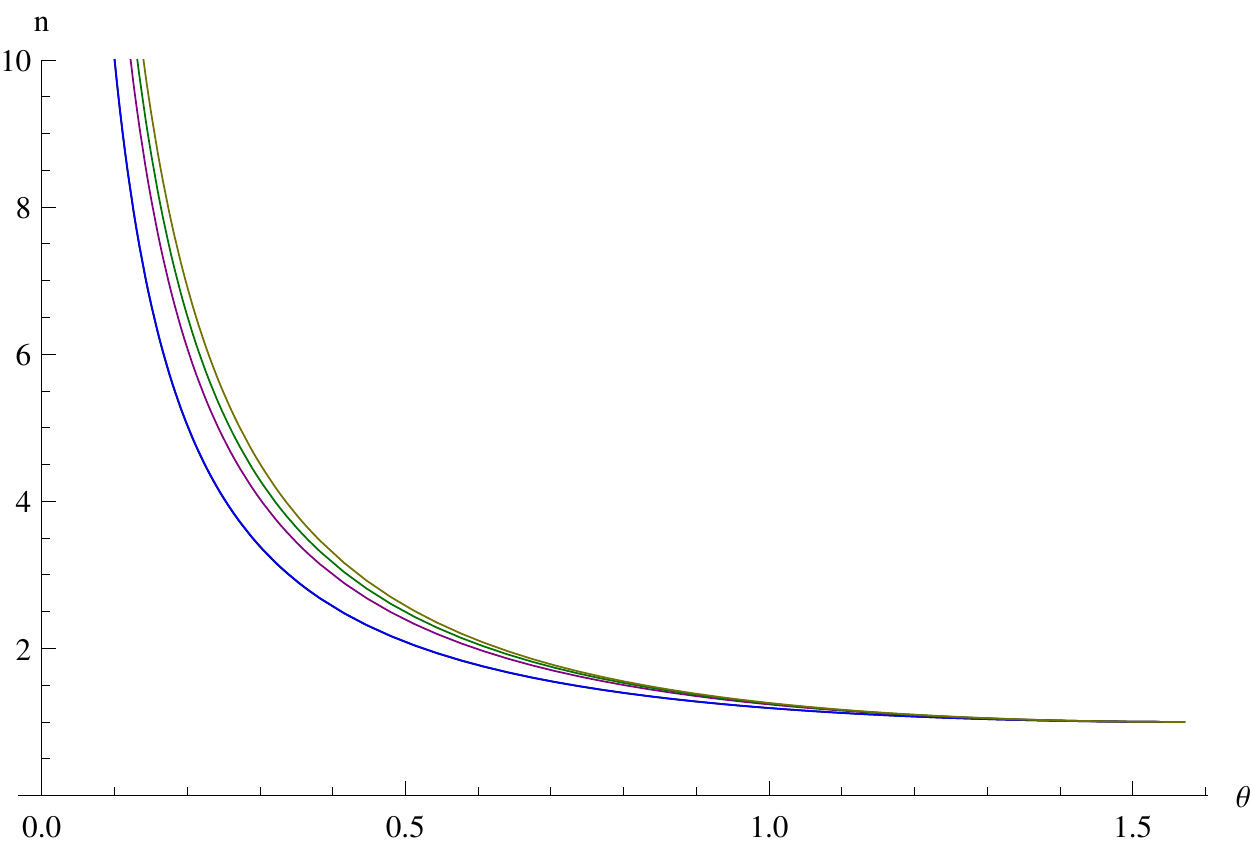}
\hskip 2cm
\includegraphics[width=6 cm, height=5 cm]{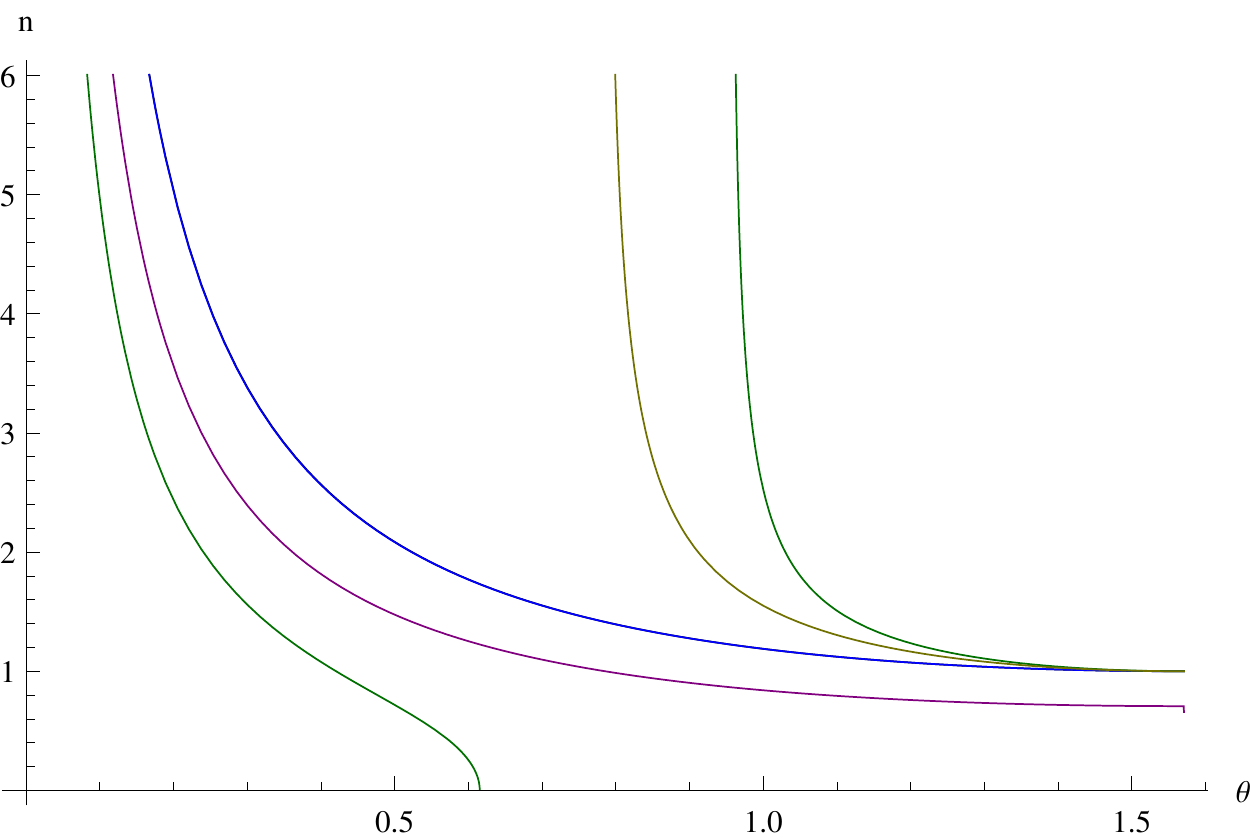}
\end{center}
\caption{The 2 types of real solutions of the light-cone equation
 for $A^\mu_\parallel$
 as a function of $\theta$ when no external $B$ is
present. On the left $n>\frac{1}{s_\theta}$, on the right
$n<\frac{1}{s_\theta}$. We vary $\alpha =
1/137\;(blue)$, $1\;(purple)$, $1.5\;(green)$, $2\;(yellow)$. The  black
($\simeq$ blue) curves are $1/\sin\theta$}
\label{fig:nnoBreal}
\end{figure}

(\ref{eq:lcnoBpar}) has seemingly
 2 types of solutions, the first with $n > \frac{1}{s_\theta}$ and
the second with $n< \frac{1}{s_\theta}$. They write respectively
\begin{equation}
\begin{split}
& \ast\ n> \frac{1}{s_\theta} :
n^2= \frac{1}{s_\theta^2}\left(1+\frac{\alpha c_\theta^2}{2(1+\alpha
s_\theta^2)}\right),\cr
& \ast\ n< \frac{1}{s_\theta} :
n^2=\frac{1}{s^2_\theta}\left(1-\frac{\alpha c^2_\theta}{2(1-\alpha
s^2_\theta)}\right).
\end{split}
\label{eq:nnoB}
\end{equation}
They are plotted on Fig.\ref{fig:nnoBreal}, respectively on the left for
$n>\frac{1}{s_\theta}$ and on the right for $n<\frac{1}{s_\theta}$. They
only depend on $\alpha$ and we plot them for $\alpha=\frac{1}{137}$ (blue),
$\alpha=1$ (purple) and $\alpha=1.5$ (green), $\alpha=2$ (yellow) together with
$n=\frac{1}{s_\theta}$ (black), the latter being in practice
indistinguishable from $\alpha=\frac{1}{137}$.

Looking at the curves, it is conspicuous that one cannot trust them,
neither when $\theta \to 0$ because of divergences, nor when $\theta$
becomes large for $\alpha >1$.
In particular, the solution with $n<\frac{1}{s_\theta}$ becomes out
of control above $\theta \geq .5$; it furthermore cannot exist
when $s^2_\theta >\frac{1}{\alpha}$ since then $n > \frac{1}{s_\theta}$ (a
divergence occurs at $s^2_\theta = \frac{1}{\alpha}$).

The approximation of considering $n\in {\mathbb R}$ is obviously
very hazardous, specially when $\alpha > 1$.
This is why we shall perform in subsection \ref{subsec:compnoB}
a detailed study with $n\in {\mathbb C}$.

\subsection{Solutions with $\boldsymbol{n\in{\mathbb C}}$}
\label{subsec:compnoB}
%=========================================================

\subsubsection{There is no
 solution with  $\boldsymbol{n_1<\frac{1}{s_\theta}}$}
\label{subsub:limtetabig}
%------------------------------------------------------

When supposing  $n\in {\mathbb R}$, we have seen that the solution with $n<
\frac{1}{s_\theta}$ was unstable, in particular above $\theta_{max}$ such
that $(\sin\theta_{max})^2 =\frac{1}{\alpha}$ were it did not exist
anymore.

Careful investigations for $n\in {\mathbb C}$ show that, like in the
presence of $B$, no solution with $n_1 <\frac{1}{s_\theta}$ exists
\footnote{In this case,
the expansion of the transmittance $V$ at small $\eta$ and $n_2$  writes
\begin{equation}
\begin{split}
\frac{1}{\pi}\Re(V) &= \frac{u n_1 c_\theta }{\sqrt{1-n_1^2
s^2_\theta}}\eta^2 + \frac12 (1+u^2)\eta^2 -\frac{n_1 s_\theta^2}{(1-n_2^2
s^2_\theta)^{\frac32}}\eta n_2 + \ldots\cr
\frac{1}{\pi}\Im(V) &=
-\frac{\eta}{\sqrt{1-n_1^2s^2_\theta}}+\frac{uc_\theta(2n_1^2
s^2_\theta-1)}{(1-n_2^2 s^2_\theta)^{\frac32}}\eta^2 n_2 +\ldots
\end{split}
\label{eq:Vcomp2}
\end{equation}
that we plug into the light-cone equation (\ref{eq:lcnoB}) for
$A^\mu_\parallel$.}.

\subsubsection{The solution with  $\boldsymbol{n_1>\frac{1}{s_\theta}}$}
\label{subsub:limtetanul2}
%---------------------------------------------------------------------

In the presence of an external $B$, we have seen that the solution with a
quasi-real index suddenly disappears below an angle $\theta_{min}\approx
\frac{1}{\Upsilon}$. In the present case with no external $B$, there is no
$\theta_{min}$ but the index becomes ``more and more complex'' (that is the
ratio of its imaginary and real parts increase) when $\theta$ becomes
smaller and smaller.

To demonstrate this, we study the light-cone equation (\ref{eq:lcnoB}) for
$A^\mu_\parallel$ with $n=n_1 + in_2,\ n_1,n_2 \in{\mathbb R}$. For
practical reasons, we shall limit ourselves to the expansion of $V$ at
small $\eta$ and $n_2$, valid when the 2 poles of $V$ lie in different 1/2
planes, given in (\ref{eq:Vexpcomp}). 

%CCCCCCCCCCCCCCCCCCCCCCCCCCCCCCCCCCCCCCCCCCCCCCCCCCCCCCCC
\begin{comment}
Calling
\begin{equation}
c= (n_1^2-n_2^2)s^2_\theta-1, \quad d= -2n_1n_2 s^2_\theta,
\end{equation}
the non-trivial parts of the
 real and imaginary parts of this light-cone equation write (we have
chosen the appropriate sign in (\ref{eq:lcnoB}, the one that gives
solutions)
\begin{equation}
\begin{split}
& 1+\frac{\alpha}{\eta\pi}\left[
\left(-\frac{c^2_\theta}{2}\frac{d}{\sqrt{2}\sqrt{c^2+d^2}\sqrt{-c+\sqrt{c^2+d^2}}}
 +s^2_\theta\frac{-d}{\sqrt{2}\sqrt{-c+\sqrt{c^2+d^2}}} \right)
\Re(V)\right . \cr
& \hskip 1cm \left .-\left(-\frac{c^2_\theta}{2}\frac{\sqrt{-c+\sqrt{c^2+d^2}}}{\sqrt{2}\sqrt{c^2+d^2}}
 + s^2_\theta\frac{\sqrt{-c+\sqrt{c^2+d^2}}}{\sqrt{2}} \right) \Im(V)
\right]=0,\cr
&
\left(-\frac{c^2_\theta}{2}\frac{d}{\sqrt{2}\sqrt{c^2+d^2}\sqrt{-c+\sqrt{c^2+d^2}}}
+s^2_\theta\frac{-d}{\sqrt{2}\sqrt{-c+\sqrt{c^2+d^2}}} \right) \Im(V) \cr
&  \hskip 1cm +
\left(-\frac{c^2_\theta}{2}\frac{\sqrt{-c+\sqrt{c^2+d^2}}}{\sqrt{2}\sqrt{c^2+d^2}}
 + s^2_\theta\frac{\sqrt{-c+\sqrt{c^2+d^2}}}{\sqrt{2}} \right) \Re(V)=0.
\end{split}
\end{equation}
\end{comment}
%CCCCCCCCCCCCCCCCCCCCCCCCCCCCCCCCCCCCCCCCCCCCCCCCCCCCC

The results are displayed in Fig.\ref{fig:nnoBcomp} below, for $\alpha=1$ (blue),
$\alpha=1.5$ (purple) and $\alpha=2$ (green). The values of $n_1$ are
plotted on the left and the ones of $n_2$ on the right. The value of the
other parameters are $u=.5,\eta=\frac{5}{1000}$.
For $\alpha=\frac{1}{137}$, $n_1$ is indistinguishable from
$\frac{1}{s_\theta}$.

\begin{figure}[h]
\begin{center}
\includegraphics[width=6 cm, height=5 cm]{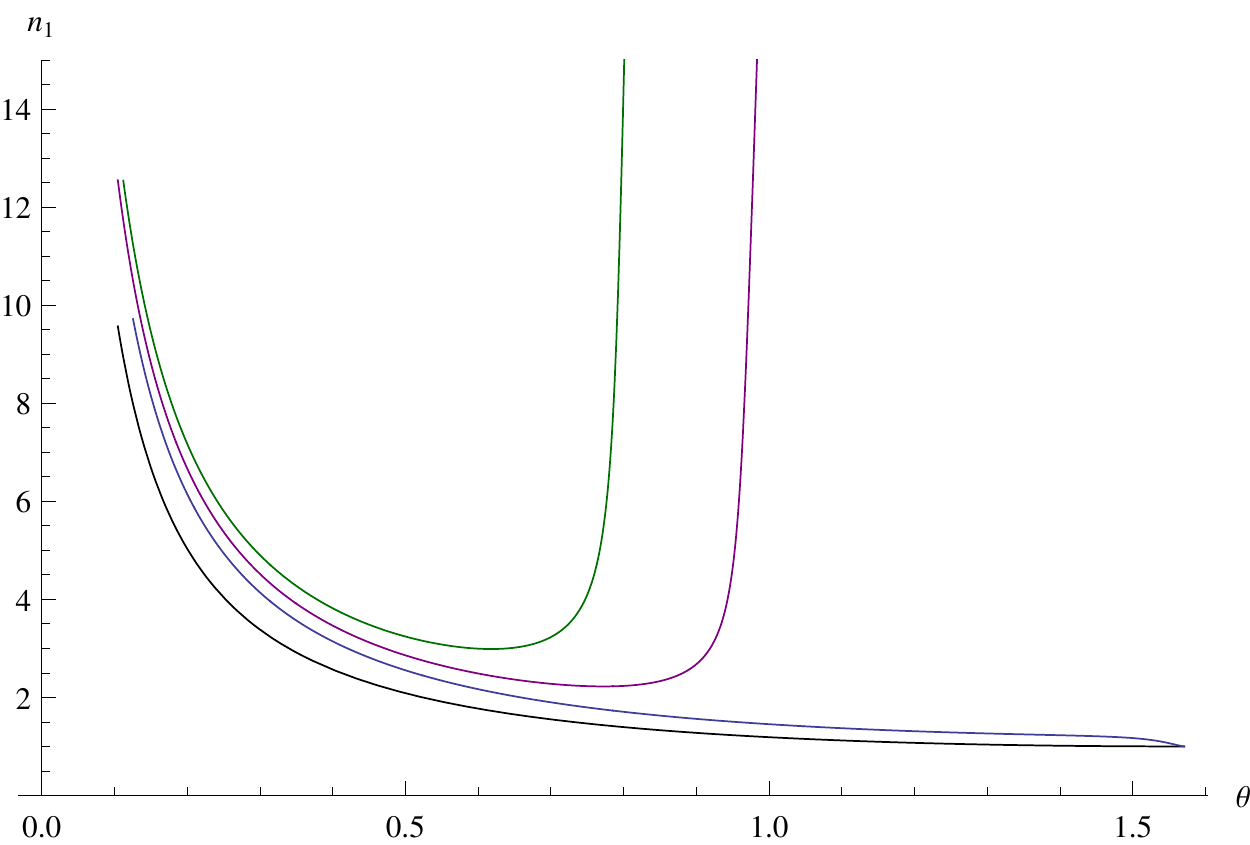}
\hskip 2cm
\includegraphics[width=6 cm, height=5 cm]{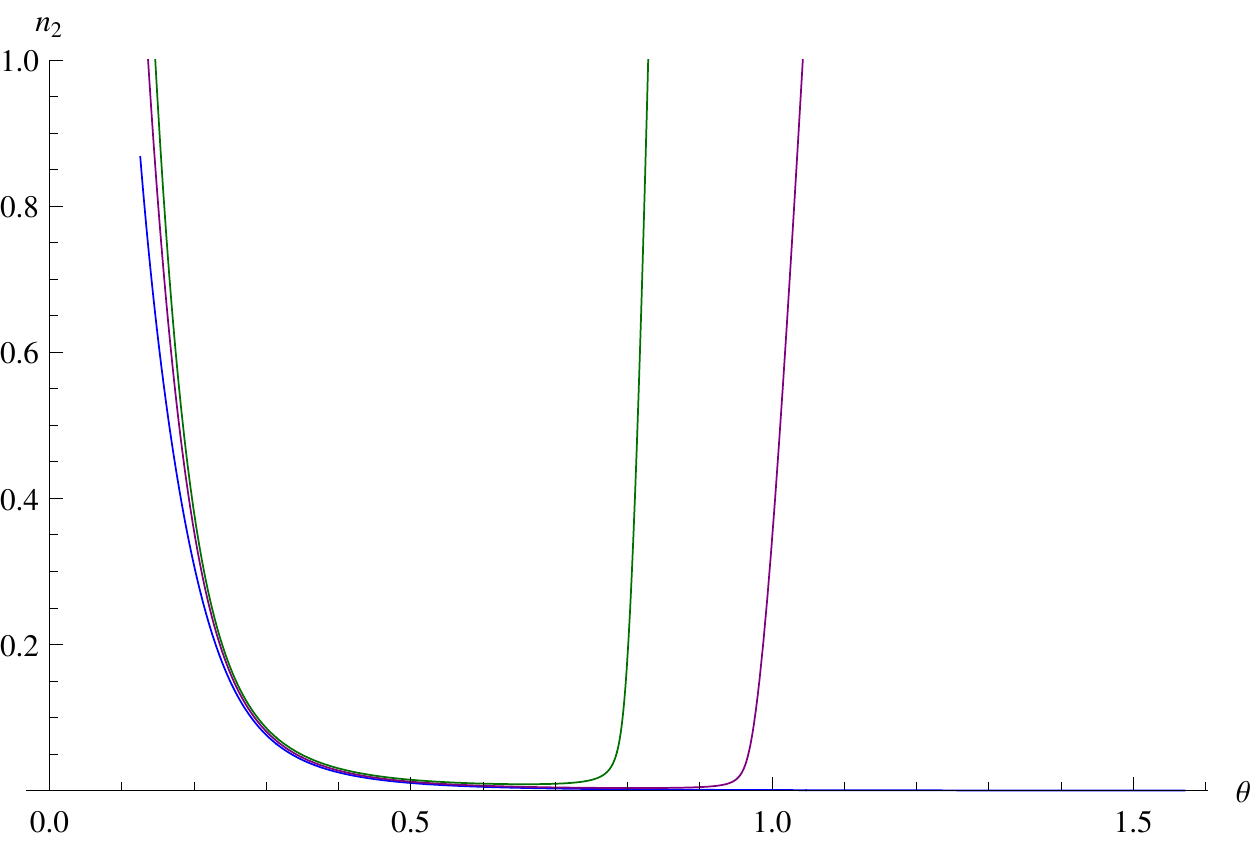}
\end{center}
\caption{The real part $n_1$ (left) and imaginary part $n_2$ (right) of the
solution $n$ of the light-cone equation (\ref{eq:lcnoB}) for
$A^\mu_\parallel$
in the absence of external $B$. The blue curves correspond to $\alpha=1$,
the purple curves to $\alpha=1.5$ and the green curves to $\alpha=2$. The
black curve on the left is $n_1=\frac{1}{s_\theta}$}
\label{fig:nnoBcomp}
\end{figure}

$\bullet$\ As $\theta$ gets smaller and smaller, the index
becomes complex with larger and larger values of both its components.
It is of course bounded as before to $|n| < \frac{1}{\eta}$ by quantum
considerations. The brutal transition at
 $\theta_{min} \simeq \frac{1}{\Upsilon}$  is replaced by a smooth
transition (which could be anticipated since, in the absence of $B$,
 the parameter $\Upsilon$ does not exist).

$\bullet$\ A new feature seems  to occur, the presence of a ``wall'' at large
$\theta$ for $\alpha > 1$,
 obviously reminiscent of the divergence that occurred in the
approximation $n\in {\mathbb R}$ at $\theta = \theta_{max},
(\sin\theta_{max})^2 = \frac{1}{\alpha}$ for the solution $n<
\frac{1}{s_\theta}$ (we had noticed that this condition could no
longer be satisfied since, for $s^2_\theta > \frac{1}{\alpha}$, $n$ could
only be larger than $\frac{1}{s_\theta}$).

Three explanations come to the mind concerning this wall. The first is that,
for large values of $n_2$, the expansion (\ref{eq:Vexpcomp})
 that we used for $V$ is no longer valid; however, using the exact expression for
the transmittance leads to the same conclusion. The second,
 and also very likely one,
is  that the perturbative series has no meaning whatsoever for
$\alpha >1$ (2-loop corrections become larger than 1-loop etc ); using
a 2-loop calculation of the vacuum polarization without external $B$
 seems feasible but also goes beyond the scope of this work.
The third is that this divergence is the sign that some
 physical phenomenon occurs, like total reflexion,  for $\theta > \theta_{max}$,
which can only be settled by experiment.

$\bullet$\ These calculations show in which domain the approximation $n \in
{\mathbb R}$ is reliable since it requires $n_2 \ll 1$: for example $n_2 <
.1$ needs $.3 \leq \theta \leq \theta_{max}$, which leaves (except for
$\alpha \leq 1$ in which case $\theta_{max} \geq \frac{\pi}{2}$) only a
small domain for $\theta$.

\paragraph{$\bullet$\ A very weak dependence on $\boldsymbol \alpha$ for
$\boldsymbol{\theta < \theta_{max}}$} 
%-----------------------------------------------------------

{\ }

In the absence of external $B$ and away from the ``wall'' at large $\theta$,
 the index is seen to depend very little on
$\alpha$.
The dependence of $n$ on $\theta$ is practically only due to the
transmittance function $V$ and to the confinement of electrons inside
graphene. Notice in particular that, when $\alpha=\frac{1}{137} \ll 1$, the
curve is indistinguishable from that of $\frac{1}{s_\theta}$.

The fairly large dependence on $\alpha$  that we uncovered
in the presence of $B$ are therefore triggered by $B$ itself.

\paragraph{$\bullet$\ The dependence on the energy of the photon}
%----------------------------------------------------------------

\begin{figure}[h]
\begin{center}
\includegraphics[width=6 cm, height=5 cm]{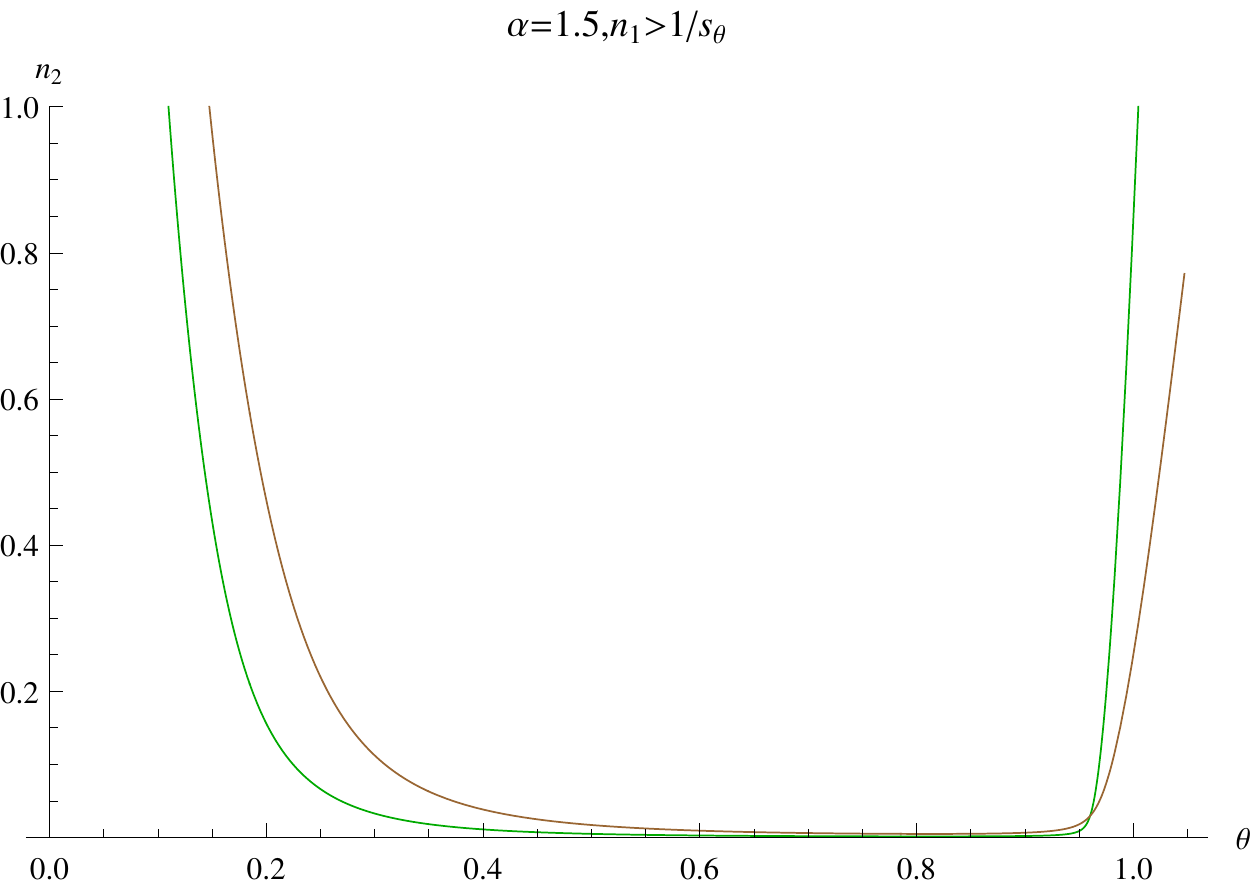}
\end{center}
\caption{$n_2$ as a function of $\theta$ for $\eta=\frac{2}{1000}$
(green) and $\eta=\frac{7}{1000}$ (brown), in the case $\alpha=1.5$}
\label{fig:n2noB}
\end{figure}

{\ }
The dependence on $\eta$ only occurs in the imaginary part $n_2$ of $n$.
This is shown in Fig.\ref{fig:n2noB}, in which we vary $\eta$ in the visible spectrum,
$\eta \in [\frac{2}{1000}, \frac{7}{1000}]$ at $\alpha=1.5$ (unlike in
Fig.\ref{fig:nnoBcomp}, $\theta$ has not been extended above $\theta_{max}$).

\subsection{The limit of very small $\boldsymbol \theta$; absorption of
visible light and  experimental opacity}
%======================================================================

\subsubsection{At small $\boldsymbol \theta$}
%--------------------------------------------

Since absorption of visible light by graphene
 at close to normal incidence has been measured \cite{Nair},
let us show that our simple model
gives predictions that are compatible with these measurements.

To that purpose, we calculated numerically the index $n$ at
the lowest value of $\theta$ at which  the
2 poles of $V$  lie in different 1/2 planes. We have used the exact
expression of $V$ (no expansion) and obtained
\begin{equation}
\begin{split}
\ast\ & for\ \alpha=1\  and\ \theta = \frac{\pi}{105.9} : n= 41.20 +
.7\times i,\cr
\ast\ & for\ \alpha=2\ and\ \theta = \frac{\pi}{89} : n = 40 + 1 \times
i.
\end{split}
\end{equation}
The 2 corresponding angles are small enough to be considered close to
normal incidence.

The transmission coefficient along $z$ (therefore for $c_\theta \approx 1$)
 is given by
\begin{equation}
T = e^{-8\pi\eta n_2 c_\theta} \approx 1- 8\pi \eta n_2,
\end{equation}
while experimental measurements \cite{Nair} are compatible with 
\begin{equation}
T \approx 1 - \pi\,\alpha_{vac},\ \alpha_{vac}=\frac{1}{137}.
\end{equation}
This requires
\begin{equation}
n_2 \approx \frac{\alpha_{vac}}{8\eta} \in [.46,.13]\ for\ \eta \in
\left[\frac{2}{1000}, \frac{7}{1000}\right].
\end{equation}

We get therefore the correct order of magnitude for $n_2$. The discrepancy
between our prediction and the experimental value can be thought as an
estimate of 2-loop corrections in the absence of $B$.

\subsubsection{At $\boldsymbol{\theta=0}$}
%========================================

 At $\theta=0$, $\vec q$ is parallel
to the $z$ axis such that there is no more distinction
between  transverse and parallel  polarizations.
Since we have found for $\theta \not=0$
 no non-trivial solution to the light-cone equation for
$A^\mu_\perp$, but only the trivial one $n=1$,
a smooth limit at $\theta =0$, which should be common for the 2
polarizations, would presumably require that, like for $A^\mu_\perp$, 
only  $n=1$ remains for
$A^\mu_\parallel$, too; but we have yet no proof of this.

So, like in the presence of $B$, we are at a loss to give any prediction
at $\theta=0$. This is for sure a limitation of our model.

%SSSSSSSSSSSSSSSSSSSSSSSSSSSSSSSSSSSSSSSSSSSSSSSSSSSSSSSS
\section{Outlook and prospects}\label{section:conclusion}
%SSSSSSSSSSSSSSSSSSSSSSSSSSSSSSSSSSSSSSSSSSSSSSSSSSSSSSSS

\subsection{General remarks}
%===========================

We have shown that the refractive index of graphene in the presence of an
external magnetic field is very sensitive to 1-loop quantum corrections.
The effects are large for optical wavelengths and even for
magnetic fields below 20 Teslas. They only depend (at least for the real
part of the refractive index), on the ration
$\frac{\sqrt{2eB}}{q_0}$ which makes them larger and larger as
the photon goes to smaller and smaller energy. We only found them for
so-called parallel polarization of the photon.
At the opposite,  when there is no external $B$,  quantum effects stay
small and the optical properties of graphene are mainly controlled by the
sole transmittance function which incorporates the geometry of the sample and
the confinement of electrons along $z$.

By calculating the 1-loop photon propagator in position space, we have been
able to localize the interactions of photons with electrons inside graphene,
therefore accounting for their ``confinement'' inside a very thin strip.

One of the main achievements of this study concerns the
transmittance function $U$. The optical properties of graphene cannot be
indeed deduced  by the sole calculation of the genuine vacuum
polarization, would it be in ``reduced $QED_{4,3}$'' \cite{KotikovTeber},
 because this would  in particular
 neglect all effects due to  the confinement of  the electron-photon
interactions.

The behavior of the refraction index as $\theta$ goes to small values has
been shown to  depend whether an external $B$ is
present or not. When $B\not= 0$ there exists a brutal transition at
$\theta_{min} \approx \frac{1}{\Upsilon}$ below which the quasi-real
solution valid above this threshold disappears, presumably (but this is
still to be proved rigorously) in favor of a complex solution with large
values of $n_1$ and $n_2$ (see subsection \ref{subsec:smallangle}).
 In the absence of $B$ the transition is smooth:
$n$ becomes gradually complex with larger and larger values of its real and
imaginary components.

Because of the approximations that we have made, and that we list below, we
cannot pretend to have devised a fully realistic quantum model.
We have indeed:\newline
*\ truncated the perturbative series at 1-loop;\newline
*\ truncated  the expansion of the electron propagator for large $B$
 at next-to-leading order;\newline
*\ approximated an incomplete $\beta$ function $F(x)= (-2)^{(-1+x)} \beta(-2,1-x,0)
\approx \frac{1}{1-x}$, which in particular forget about poles at $p_0^2=
2neB$ except for $n=1$; this is however safe for electrons with energy lower than
$7\, eV$, which is certainly the case for graphene through which go photons
with energies smaller than $3.5\,eV$;\newline
* chosen a special gauge, the Feynman gauge for the external
photons;\newline
*\ studied light-cone equations only through their expansions at large
$\frac{\sqrt{2eB}}{q_0}$ and small $aq_0$.

We can however  reasonably pretend
to  have gone beyond the brutal limit $B\to \infty$ and
to have defined  a domain of wavelengths and magnetic fields in which
specific expansions and approximations are under control and which are
furthermore physically easy to test.

Some comments are due concerning  the lack of transversality of the
vacuum polarization which arises here, as well as in \cite{Godunov}, from the
interaction of ``quasi-2+1'' electrons (in reality 3+1 electrons with $p_3$
formally vanishing) with 3+1 photons. Lorentz invariance being explicitly
 broken, one cannot expect anymore the usual gauge invariance of 3+1 QED to
hold like in \cite{TsaiErber1}. 

A specific choice of gauge appears then less chocking, all the more as
it is extremely common when making
calculations in condensed matter physics to choose the most convenient
(Coulomb or Feynman) gauge.

A tantalizing question concerns of course the magnitude of higher order
corrections. If 1-loop corrections to the refraction index are large, how
can we trust the result, unless  all higher orders are proved to be
much smaller? At present we have no answer to this.
That $\alpha \simeq 2$
inside graphene is already a  bad ingredient for a reliable
 perturbative treatment
\footnote{In the case of the hydrogen atom it was shown in \cite{Godunov}
that 2-loop effects are negligible. It is also instructive to look at
\cite{HofmannBarnesSarma} which show that, in the framework of the Random Phase
Approximation and making a 2-loop calculations, graphene, despite
a large value of $\alpha$, behaves like a weakly coupled system.
However, in this study, no external magnetic field is present.}
and,
furthermore, the corrections to $n$ do not look like a standard
series in powers of $\alpha$. Comparisons can be made for example with the
results obtained in the case of non-confined massive electrons with
the effective Euler-Heisenberg Lagrangian
\cite{DittrichGies1}. Their equations (2.17)(2.18) show quantum corrections
to $n$ proportional to $\alpha\left(\frac{e B}{m_e^2}\right)^2$.
 In the study of the
hydrogen atom \cite{MachetVysotsky}\cite{GodunovMachetVysotsky}, typical
corrections are proportional to $\alpha \frac{eB}{m_e^2}$. In the present
study, electrons are massless, and dimensionless factors are built
with $q_0$ in place of $m_e$. Quantum corrections to the leading
$\frac{1}{s_\theta}$ behavior of the index come out proportional to
$\left(\frac{\alpha}{\pi}\right)^2\frac{eB}{q_0^2}$ (see (\ref{eq:solpar})),
 which is  very unusual.

\subsection{Going below $\boldsymbol{\theta_{min}}$ in the presence of
$\boldsymbol B$; are there also solutions with a large absorption?}
\label{subsec:smallangle}
%====================================================================

We have seen that, as $\theta$ decreases, a transition occurs at $\theta
\sim \frac{1}{\Upsilon} = \frac{q_0}{\sqrt{2eB}}$. The quasi-real solution
that we have exhibited for larger angles disappears.

\begin{figure}[h]
\begin{center}
\includegraphics[width=6 cm, height=5 cm]{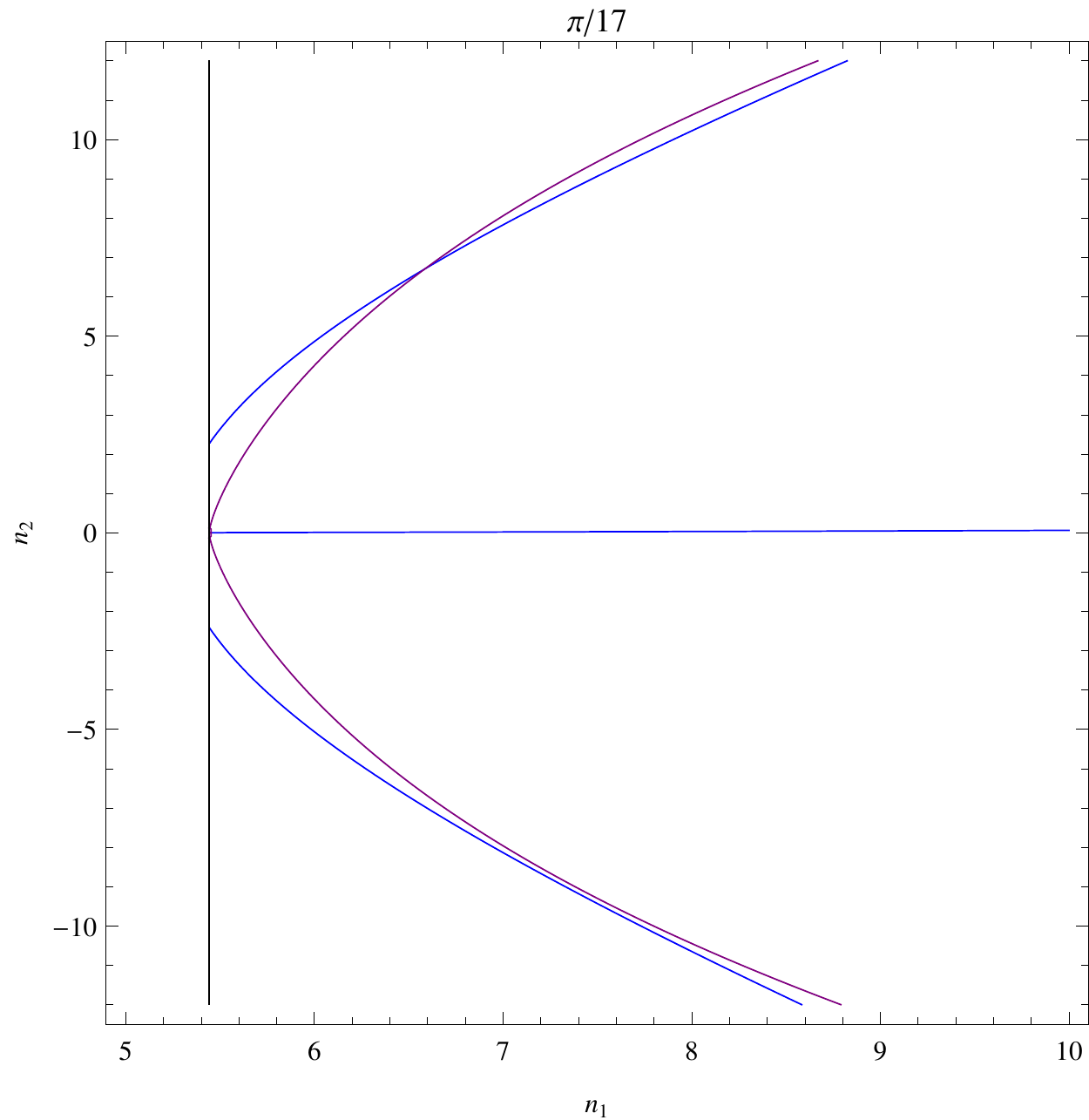}
\end{center}
\caption{Solutions of the real part (purple) and imaginary part (blue)
of the light-cone equation (\ref{eq:lccomp}) for $A^\mu_\parallel$ at
$\theta= \frac{\pi}{17}$ in the presence of $B$.
 The black vertical line on the left corresponds to $n_1=\frac{1}{s_\theta}$}
\label{fig:othersol}
\end{figure}

If one considers, below the threshold, at the same $\theta=\frac{\pi}{17}$
the same Fig.\ref{fig:wall1} drawn on a much larger domain for $n_1$ and $n_2$, one gets
Fig.\ref{fig:othersol}. One solution (at least) occurs for the light-cone equation
(\ref{eq:lccomp}),
which corresponds to $n_1 \approx 6.5, n_2 \approx 7$.

This suggests that below $\theta_{min}$, the system goes to a large index
with a large absorption. This type of solution is incompatible with the
 approximations that we have made to find them $|n_2| \ll n_1$ etc, such that
drawing a definitive conclusion requires using more elaborate numerical
methods. This will be the subject of a forthcoming work.
Let us only mention here in addition that such solutions with large
index/absorption may coexist, above $\theta_{min}$, with the quasi-real
solutions that we have exhibited in this work.

Whether or not total reflection occurs inside graphene at
low incidences can only be settled with
this more complete study. Notice that, if such a
phenomenon occurs, it is at small angle of incidence,
 again at the opposite of what one is accustomed to
with geometrical optics.

A brutal transition like this one may also be the sign of a phase
transition at the level of fermions or photons. It has often been evoked
that chiral symmetry may get broken inside graphene in the presence of a
magnetic field (see for example \cite{chiral}), and that the photon eventually gets
an effective mass (breaking of gauge invariance)
should also not be systematically rejected before careful investigations
have been done.

\subsection{A bridge between Quantum Field Theory, quantum optics and
nanophysics}
%===================================================================

Along this limited study, we have pointed at other potentially interesting
phenomena that deserve more detailed investigations: a brutal transition
below $\theta=\theta_{min}$ in the presence of $B$, the eventual
existence, in the same conditions,
 of several types of solutions (including some with large $n_1$
and $n_2$), the presence
of a ``limiting $B=B^m$''  above which new quantum effects are expected,
and, even in the absence of $B$, some intriguing behavior of the refractive
index above $\theta=\theta_{max}$  for $\alpha >1$.

The issue whether graphene can be safely
described in perturbation theory despite a large electromagnetic coupling
deserves also, of course, deeper investigations.

The wavelengths of visible light are $\sim 1000$ times larger than the
thickness of graphene. The laws of refraction are therefore not expected
to be true.
This is confirmed by the existence of solutions to the light-cone equations
only satisfying the condition $n \sin\theta >1$,  $n$ being the index
inside graphene. Since  $\theta$ has also been defined as the angle of incidence
inside the medium, it is manifestly impossible to satisfy the laws of
refraction at its interface with vacuum, which would write $n \sin\theta
 = 1 \times \sin\theta_{vacuum}$: the l.h.s. is indeed  $>1$ while the
 r.h.s. is  $\leq 1$.

Graphene in external magnetic field is thus certainly not the realm
 of geometrical optics, but  it could well prove, inversely, a privileged
test-ground for the interplay between Quantum Field Theory,
 quantum optics and nanophysics.

%-------------------------------------------------------------------------

\vskip .5cm

{\em \underline{Acknowledgments:} it is a pleasure to thank M.~Vysotsky
for his continuous interest and  encouragements.}

%%%%%%%%%%%%%%%%%%%%%%%%%%%%%%%%%%%%%%%%%%%%%%%%%%%%%%%%%%%%%%%%%%%%%%%%%%
\newpage

\begin{em}

\end{em}

%=======================================================================
\end{document}